\documentclass[10pt,journal,compsoc]{IEEEtran}
\usepackage{tabularx}
\usepackage{graphicx}
\usepackage{url}
\usepackage{subcaption}
\usepackage{helvet} 
\usepackage{courier} 
\usepackage{hyperref}
\usepackage{amssymb}
\usepackage{amsthm}
\usepackage{array}
\usepackage{mdwlist}
\usepackage{amsmath}
\usepackage{float}
\usepackage[labelfont=bf, textfont=bf]{caption}
\usepackage[linesnumbered,ruled]{algorithm2e}
\usepackage{algorithmic}
\usepackage{enumitem}
\setlist{leftmargin=5mm}

\newcolumntype{C}[1]{>{\centering\let\newline\\\arraybackslash\hspace{0pt}}m{#1}}
\newcommand{\ie}{\emph{i.e., }}
\newcommand{\eg}{\emph{e.g., }}
\newcommand{\etal}{\emph{et al.}}

\newcommand{\wrt}{\emph{w.r.t. }}
\newcommand{\cf}{\emph{cf. }}

\newcommand{\aka}{\emph{aka. }}

\begin{document}
\title{NAIS: Neural Attentive Item Similarity Model for Recommendation}
\author{Xiangnan~He, Zhankui~He, Jingkuan~Song, Zhenguang~Liu, Yu-Gang Jiang and Tat-Seng Chua
\IEEEcompsocitemizethanks{
	\IEEEcompsocthanksitem X. He and T.-S. Chua are with School of Computing, National University of Singapore, Singapore. E-Mail: xiangnanhe@gmail.com and dcscts@nus.edu.sg
	\IEEEcompsocthanksitem Z. He and Y.-G. Jiang are with School of Computer
	Science, Fudan University, China. Email: zkhe15@fudan.edu.cn and ygj@fudan.edu.cn
	\IEEEcompsocthanksitem J. Song is with University of Electronic Science and Technology of China, China. Email: jingkuan.song@gmail.com
	\IEEEcompsocthanksitem Z. Liu is with A* STAR, Singapore. Email: liuzhenguang2008@gmail.com
	}
\thanks{Zhenguang Liu is the corresponding author.}
}

\markboth{IEEE Transactions on Knowledge and Data Engineering, 2018}
{}

\IEEEtitleabstractindextext{
\begin{abstract}
Item-to-item collaborative filtering (\aka item-based CF) has been long used for building recommender systems in industrial settings, owing to its interpretability and efficiency in real-time personalization. It builds a user's profile as her historically interacted items, recommending new items that are similar to the user's profile. As such, the key to an item-based CF method is in the estimation of item similarities. Early approaches use statistical measures such as cosine similarity and Pearson coefficient to estimate item similarities, which are less accurate since they lack tailored optimization for the recommendation task. In recent years, several works attempt to learn item similarities from data, by expressing the similarity as an underlying model and estimating model parameters by optimizing a recommendation-aware objective function. While extensive efforts have been made to use shallow linear models for learning item similarities, there has been relatively less work exploring nonlinear neural network models for item-based CF. 

\quad In this work, we propose a neural network model named \textit{Neural Attentive Item Similarity model} (NAIS) for item-based CF. The key to our design of NAIS is an attention network, which is capable of distinguishing which historical items in a user profile are more important for a prediction. Compared to the state-of-the-art item-based CF method \textit{Factored Item Similarity Model} (FISM)~\cite{FISM}, our NAIS has stronger representation power with only a few additional parameters brought by the attention network. Extensive experiments on two public benchmarks demonstrate the effectiveness of NAIS. This work is the first attempt that designs neural network models for item-based CF, opening up new research possibilities for future developments of neural recommender systems. 
\end{abstract}

}

\maketitle

\IEEEdisplaynotcompsoctitleabstractindextext
\IEEEpeerreviewmaketitle

\section{Introduction}
\label{sec:introduction}
\IEEEPARstart
Recommender system is a core service for many customer-oriented online services to increase their traffic and make profits, such as E-commerce and social media sites. For example, it was reported that in YouTube, recommendations accounted for about $60\%$ video clicks for the homepage~\cite{youtube2010}; in Netflix, recommender systems contributed about $80\%$ of movies watched and placed the business value of over \$1 billion per year, as indicated by their Chief Product Officer Neil Hunt~\cite{netflix2015}. 

In modern recommender systems, collaborative filtering (CF) --- a technique that predicts users' personalized preference from user-item interactions only --- plays a central role especially in the phase of candidate generation~\cite{liu2017related,NCF}. 
Popularized by the Netflix Prize, matrix factorization (MF) methods have become the most popular recommendation approach in academia and been widely studied in literatures~\cite{fastMF,DCF}. While MF methods are shown to provide superior accuracy over neighbor-based methods in terms of rating prediction, they have been relatively seldom reported to be used in industrial applications. 
One possible reason is due to MF's personalization scheme --- user-to-item CF that characterizes a user with an ID and associates it with an embedding vector. As a result, to refresh recommendations for a user with her new interactions, the user's embedding vector has to be updated. However, re-training a MF model for large-scale data is difficult to achieve in real time and may require complex software stack to support online learning, making the approach less attractive for industrial settings~\cite{iCD}. 

On the other hand, item-to-item CF --- which characterizes a user with her historically interacted items and recommends items similar to the user's profile --- has been heavily used in industrial applications~\cite{liu2017related,youtube2010,netflix2015,smith2017two}. Not only does item-based CF provide more interpretable prediction suitable for many recommendation scenarios, but it also makes real-time personalization much easier to achieve. Specifically, the major computation that estimates item similarities can be done offline and the online recommendation module only needs to perform a series of lookups on similar items, which can be easily done in real-time. 

Early item-based CF approaches use statistical measures such as Pearson coefficient and cosine similarity to estimate item similarities~\cite{ICF}. Since such heuristic-based approaches lack tailored optimization for recommendation, they typically underperform machine learning-based methods in terms of top-K recommendation accuracy~\cite{BPR,fastMF}. 
To tackle this, Ning \etal~\cite{SLIM} adopt a machine learning view for item-based CF, which learns item similarity from data by optimizing a recommendation-aware objective function. 
Although better accuracy can be achieved, directly learning the whole item--item similarity matrix has a quadratic complexity \wrt the number of items, making it infeasible for practical recommenders that need to deal with millions or even billions of items. 

To address the inefficiency issue of learning-based item-to-item CF, Kabbur \etal~\cite{FISM} propose a factored item similarity model (FISM), which represents an item as an embedding vector and models the similarity between two items as the inner product of their embedding vectors. Being a germ of representation learning~\cite{representation_learning,luo2016image}, FISM provides state-of-the-art recommendation accuracy and is well suited for online recommendation scenarios. However, we argue that FISM's modeling fidelity can be limited by its assumption that all historical items of a user profile contribute equally in estimating the similarity between the user profile and a target item. Intuitively, a user interacts with multiple items in the past, but it may not be true that these interacted items reflect the user's interest to the same degree. For example, a fan of affectional films might also watch a horror film just because the film was popular during that time. Another example is that user interests may change with time, and as such, recently interacted items should be more reflective of a user's future preference. 

In this work, we propose an enhanced item similarity model by distinguishing the different importance of interacted items in contributing to a user's preference. Our NAIS model is built upon FISM, preserving the same merit with FISM in terms of high efficiency in online prediction, while being more expressive than FISM by learning the varying importance of the interacted items. This is achieved by employing the recent advance in neural representation learning --- the attention mechanism~\cite{ijcai2017-afm,ACF,Li:2017:NAS} --- for learning item-to-item interactions. 
One of our key findings is that the standard attention mechanism fails to learn from users historical data, due to the large variance on the lengths of user histories. 
To address this, we adjust the attention design by smoothing user histories.  
We conduct comprehensive experiments on two public benchmarks to evaluate top-K recommendation, demonstrating that our NAIS betters FISM for a $4.5\%$ relative improvement in terms of NDCG and achieves competitive performance. To facilitate the research community to validate and make further developments upon NAIS, we have released our implementation codes in:  \url{https://github.com/AaronHeee/Neural-Attentive-Item-Similarity-Model}. 

The remainder of the paper is as follows. After introducing some preliminaries in Section~\ref{sec:pre}, we elaborate our proposed method in Section~\ref{sec:method}. We then perform experimental evaluation in Section~\ref{sec:exper}. We discuss related work in Section~\ref{sec:related}, before concluding the whole paper in Section~\ref{sec:conclusion}. 

\section{Preliminaries}
\label{sec:pre}
We first shortly recapitulate the standard item-based CF technique~\cite{ICF}. We then introduce the learning-based method for item-based CF~\cite{SLIM} and FISM~\cite{FISM}, which are building blocks for our proposed NAIS method. 

\subsection{Standard Item-based CF}
The idea of item-based CF is that the prediction of a user $u$ on a target item $i$ depends on the similarity of $i$ to all items the user has interacted with in the past. Formally, the predictive model of item-based CF is:
\begin{equation} \label{eq:itemCF}
	\hat{y}_{ui} = \sum_{j\in \mathcal{R}^+_u} r_{uj} s_{ij},
\end{equation}
where $\mathcal{R}^+_u$ denotes the set of items that user $u$ has interacted with, $s_{ij}$ denotes the similarity between item $i$ and $j$, and $r_{uj}$ is an interaction denoting the known preference of user $u$ on $j$ --- for explicit feedback (\eg ratings) $r_{uj}$ can be a real value denoting the rating score, and for implicit feedback (\eg purchases) $r_{uj}$ can be a binary value 1 or 0 denoting whether $u$ has interacted with $j$ or not. 


The appealing property of efficient online recommendation is brought by its compositionality in computing the prediction score. First, when item similarities have been obtained offline, the online recommendation phase only needs to retrieve top similar items of candidate items $\mathcal{R}^+_u$ and score them with Equation (\ref{eq:itemCF}).
Second, to refresh recommendations for a user with her new interactions, we only need to consider items that are similar to the newly interacted items. This incremental complexity makes item-based CF very suitable for online learning and real-time personalization, as demonstrated in~\cite{youtube2010,iCD}. 

For the item similarity $s_{ij}$, an intuitive approach is to represent an item as its interacted users and apply similarity measures such as cosine similarity and Pearson coefficient~\cite{ICF}. Another common approach is to employ random walks on the user-item interaction graph~\cite{liu2017related}. However, such heuristic-based approaches for estimating item similarities lack optimization tailored for recommendation, and thus may yield suboptimal performance. In what follows, we introduce learning-based methods which aim to boost the accuracy of item-based CF by adaptively learning item similarities from data.

\subsection{Learning-based Methods for Item-based CF}


In \cite{SLIM}, the authors proposed a method named SLIM (short for \textit{Sparse LInear Method}), which learns item similarities by optimizing a recommendation-aware objective function. The idea is to minimize the loss between the original user-item interaction matrix and the reconstructed one from the item-based CF model. Formally, the objective function to minimize is as follows:
\begin{equation}\label{eq:SLIM}
	\begin{aligned}
		&L = \frac{1}{2}\sum_{u=1}^U \sum_{i=1}^{I} (r_{ui}-\hat{y}_{ui})^2 + \beta ||\textbf{S}||_2 + \gamma ||\textbf{S}||_1 \\
		& subject\ to \quad \textbf{S}\ge 0, \  diag(\textbf{S}) = 0,
	\end{aligned}
\end{equation}
where $U$ and $I$ denote the number of users and items, respectively, $\textbf{S}\in\mathbb{R}^{I\times I}$ denotes the item-item similarity matrix, and $\beta$ controls the strength of $L_2$ regularization for preventing overfitting. Note that in SLIM there are three purposely designed constraints on $\textbf{S}$ to ensure an effective learning of item similarities: 1) the $L_1$ regularization controlled by $\gamma$ to enforce sparsity on $\textbf{S}$, as in practice there are only a few items that are particularly similar to an item; 2) the non-negativity constraint on each element of $\textbf{S}$ to make it a meaningful similarity measure; and 3) the zero constraint on diagonal elements of $\textbf{S}$ to eliminate the impact of the target item itself in estimating the prediction. 

Despite better recommending accuracy can be achieved, SLIM has two inherent limitations. First, the offline training process can be very time consuming for large-scale data, due to the direct learning on $\textbf{S}$ that has $I^2$ elements (the time complexity is in the magnitude of $O(I^2)$). Second, it can only learn similarities for two items that have been co-rated before, and fails to capture transitive relations between items. To address the limitations, the later work \cite{FISM} proposed FISM (short for \textit{Factored Item Similarity Model}), which represents an item as a low-dimensional embedding vector; then the similarity score $s_{ij}$ is parameterized as the inner product between the embedding vector of $i$ and $j$. Formally, the predictive model of FISM is\footnote{The bias terms in the original paper are omitted for clarity.}:
\begin{equation} \label{eq:FISM}
	\hat{y}_{ui} = \textbf{p}_i^T 
	\underbrace{(\frac{1}{|\mathcal{R}^+_u|^{\alpha}}
		\sum_{j\in \mathcal{R}^+_u \setminus \{i\}}  \textbf{q}_j)}_{\text{user u's representation}},
\end{equation}
where $\alpha$ is a hyper-parameter controlling the normalization effect, $\textbf{p}_i$ and $\textbf{q}_j$ denote the embedding vector for item $i$ and $j$, respectively. The symbol $\setminus \{i\}$ corresponds to the constraint of $diag(\textbf{S}) = 0$ in Equation~(\ref{eq:SLIM}), to avoid the modeling of self-similarity of the target item. 

From the view of user-based CF, the term in the bracket can be seen as the user $u$'s representation, which is aggregated from the embeddings of historical items of $u$. 
Note that in FISM, each item has two embedding vectors $\textbf{p}$ and $\textbf{q}$ to differentiate its role of a prediction target or a historical interaction, which can also increase model expressiveness; the rating term $r_{uj}$ is omitted as FISM concerns implicit feedback where $r_{uj}=1$ for $j\in \mathcal{R}^+_u$. Given the well-defined predictive model of Equation (\ref{eq:FISM}), one can learn model parameters by optimizing standard losses for recommendation (\ie without the item similarity constraints used in SLIM), 
such as the pointwise classification loss~\cite{NCF} and pairwise regression loss~\cite{SilkRoad}.

While FISM provides state-of-the-art performance among item-based CF methods, we argue that its representation ability can be limited by its equal treatments on all historical items of a user when obtaining the user's representation. As mentioned before in introduction, this assumption is counter-intuitive for real-world data and may decrease model fidelity. Our proposed NAIS model tackles this limitation of FISM by differentiating the importance of historical items with a neural attention network. 


\section{Neural Attentive Item Similarity Model}
\label{sec:method}
In this section, we present our proposed NAIS methods. Before introducing the NAIS model, we first discuss several designs of attention mechanism that attempt to address the limitation of FISM. We then elaborate the optimization of model parameters. We focus the discussion of optimizing NAIS with implicit feedback, which is the recent focus of recommendation research since implicit feedback is more prevalent and easy to collect than explicit ratings. Lastly, we discuss several properties of NAIS, including the time complexity, support for online personalization, and options for the attention function.  

\begin{figure}[t]
	\centering
	\includegraphics[scale=0.34]{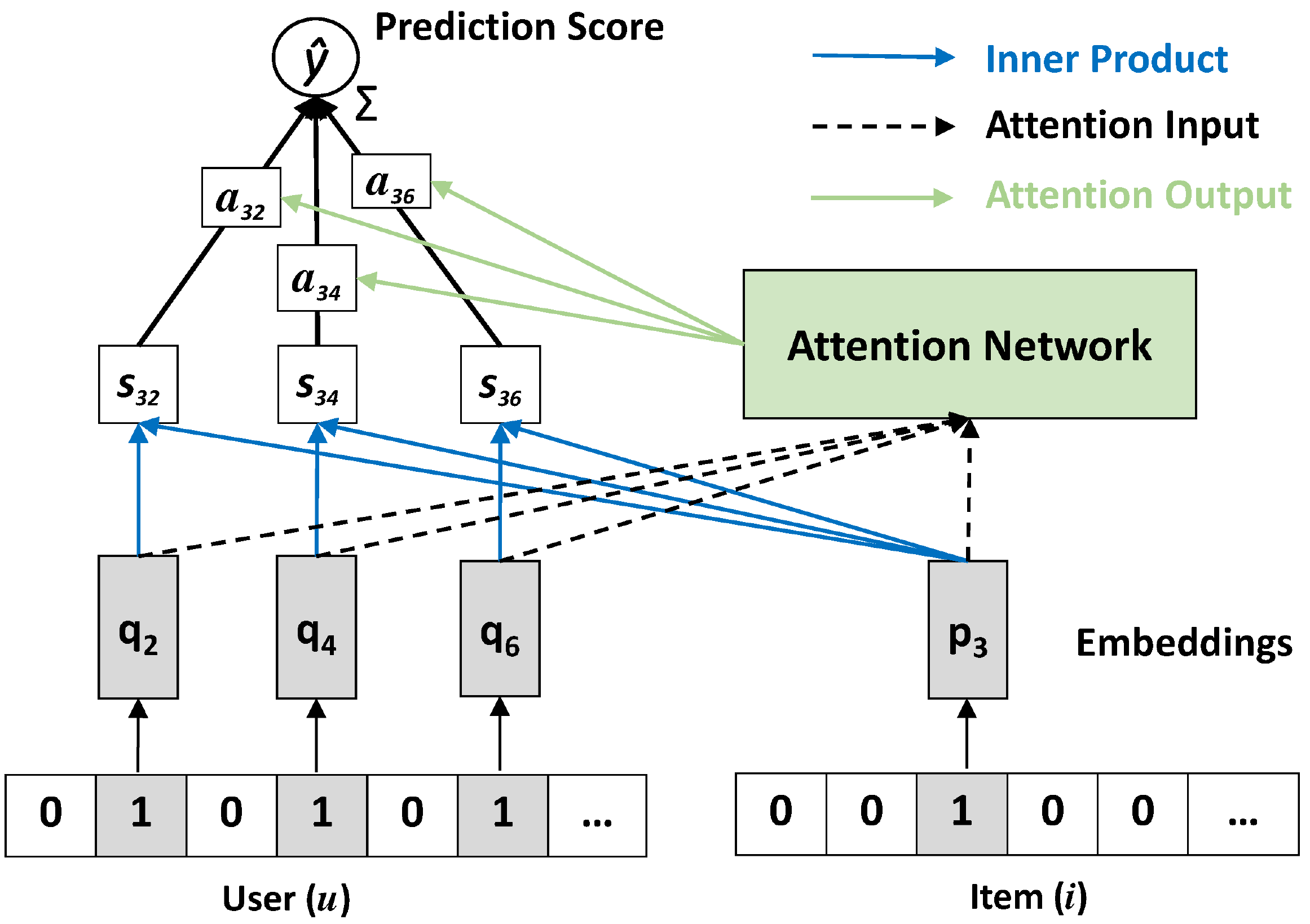}
	\vspace{-5pt}
	\caption{The neural collaborative filtering framework of our Neural Attentive Item Similarity (NAIS) model.}
	\vspace{-10pt}
	\label{fig:NAIS}
\end{figure}

\subsection{Model Designs}
\textbf{Design 1}. The original idea of attention is that different parts of a model can contribute (\ie attend) differently for the final prediction~\cite{bahdanau2014neural}. In the scenario of item-based CF, we can intuitively allow historical items contributing differently to a user's representation by assigning each item an individualized weight:
\begin{equation}\label{eq:NAIS-basic}
	\hat{y}_{ui} = \textbf{p}_i^T (\frac{1}{|\mathcal{R}^+_u|^{\alpha}} \sum_{j\in \mathcal{R}^+_u \setminus \{i\}} a_j \textbf{q}_j),
\end{equation}
where $a_j$ is a trainable parameter that denotes the attention weight of item $j$ in contributing to user representation. Clearly, this model subsumes the FISM, which can be resumed by fixing $a_j$ to 1 for all items. While this model seems to be capable of differentiating the importance of historical items, it ignores the impact of the target item on a historical item.
Particularly, we argue that it is unreasonable to assign a historical item a global weight for all predictions, regardless of which item to predict. For example, when predicting a user's preference on a romantic movie, it is undesirable to consider a horrible movie as equally important as another romantic movie. From the perspective of user representation learning, it assumes that a user has a static vector to represent her interest, which may limit the model's representation ability. \vspace{+5pt}

\noindent\textbf{Design 2}. To address the limitation of Design 1, an intuitive solution is to tweak $a_j$ to be aware of the target item, \ie assigning an individualized weight for each $(i, j)$ pair:
\begin{equation}\label{eq:NAIS-target}
	\hat{y}_{ui} = \textbf{p}_i^T (\frac{1}{|\mathcal{R}^+_u|^{\alpha}} \sum_{j\in \mathcal{R}^+_u \setminus \{i\}} a_{ij} \textbf{q}_j),
\end{equation}
where $a_{ij}$ denotes the attention weight of item $j$ in contributing to user $u$'s representation when predicting $u$'s preference on target item $i$. Although this solution seems to be technically viable, the problem is that if an item pair $(i, j)$ has never co-occurred in training data (\ie no user has interacted with both $i$ and $j$), its attention weight $a_{ij}$ cannot be estimated and will be a trivial number.  \vspace{+5pt}

\noindent\textbf{Design 3}. To solve the generalization issue of Design 2, we consider relating $a_{ij}$ with the embedding vector $\textbf{p}_i$ and $\textbf{q}_j$. The rationale is that the embedding vectors are supposed to encode the information of items, thus they can be used to determine the weight of an interaction $(i,j)$. 
Specifically, we parameterize $a_{ij}$ as a function with $\textbf{p}_i$ and $\textbf{q}_j$ as the input:
\begin{equation}
	a_{ij} = f(\textbf{p}_i, \textbf{q}_j).
\end{equation}
The advantage of this parameterization is that even though a pair $(i, j)$ has never co-occurred, as long as $\textbf{p}_i$ and $\textbf{q}_j$ have been reliably learned from data, they can still be used to estimate the attention weight $a_{ij}$ well. To achieve this goal, we need to ensure the function $f$ has strong representation power. Inspired by the recent success of using neural networks to model the attention weight~\cite{ACF,ijcai2017-afm}, we similarly use a Multi-Layer Perception (MLP) to parameterize the attention function $f$. Specifically, we consider two ways to define the attention network:
\begin{equation}\label{eq:f}
	\begin{cases} 
		&1. \ f_{concat}(\textbf{p}_i,\textbf{q}_j) = \textbf{h}^T ReLU(\textbf{W} \begin{bmatrix}\textbf{p}_i \\ \textbf{q}_j \end{bmatrix} + \textbf{b}) \\
		&2. \ f_{prod}(\textbf{p}_i,\textbf{q}_j) = \textbf{h}^T ReLU(\textbf{W} (\textbf{p}_i\odot \textbf{q}_j) + \textbf{b})
	\end{cases}
\end{equation}
where $\textbf{W}$ and $\textbf{b}$ are respectively the weight matrix and bias vector that project the input into a hidden layer, and $\textbf{h}^T$ is the vector that projects the hidden layer into an output attention weight. We term the size of hidden layer as ``\textit{attention factor}'', for which a larger value brings a stronger representation power for the attention network. We use the Rectified Linear Unit (ReLU) as the activation function for the hidden layer, which has shown to have good performance in neural attention network~\cite{ijcai2017-afm}. In later Section~\ref{ss:discussions}, we discuss the pros and cons of the two attention functions $f_{concat}$ and $f_{prod}$.

Following the standard setting of neural attention network~\cite{zhaoIJCAI17,ACF}, we can formulate the predictive model of Design 3 as follows:
\begin{equation}
	\begin{aligned}\small
		\hat{y}_{ui} &= \textbf{p}_i^T ( \sum_{j\in \mathcal{R}^+_u \setminus \{i\}} a_{ij} \textbf{q}_j), \\
		a_{ij}&=\frac{\exp{(f(\textbf{p}_i, \textbf{q}_j))}}{\sum_{j\in \mathcal{R}^+_u\setminus \{i\}} \exp{(f(\textbf{p}_i, \textbf{q}_j))}},
	\end{aligned}
\end{equation}
where the coefficient $\frac{1}{|\mathcal{R}^+_u|^{\alpha}}$ is aborted into the attention weight $a_{ij}$ without affecting the representation power, and the softmax function is used to convert the attention weights to a probabilistic distribution. Note that this is the most natural and straightforward way to employ an attention network on interaction history, which is the same as the history modeling part of the \textit{Attentive CF} model~\cite{ACF}. 

Unfortunately, we find such a standard solution of attention does not work well in practice --- it underperforms FISM significantly, even though it can generalize FISM in theory. After investigating the attention weights, we unexpectedly find the problem stems from the softmax function, a standard choice in neural attention networks. The rationale is as follows. 
In conventional usage scenarios of attention such as CV and NLP tasks, the number of attentive components does not vary much, such as words in sentences~\cite{DBLP:conf/emnlp/ParikhT0U16} and regions in images~\cite{DBLP:conf/cvpr/ChenZXNSLC17,wang2015visual}. As such, using softmax can properly normalize attention weights and in turn has a nice probabilistic explanation. However, such a scenario does not exist any more for user historical data, since the history length of users (\ie number of historical items consumed by users) can vary much. 
Qualitatively speaking, the softmax function performs $L_1$ normalization on attention weights, which may overly punish the weights of active users with a long history. 

\begin{figure}[t]
	\centering
	\begin{subfigure}[b]{0.23\textwidth}
		\centering
		\includegraphics[width=\textwidth]{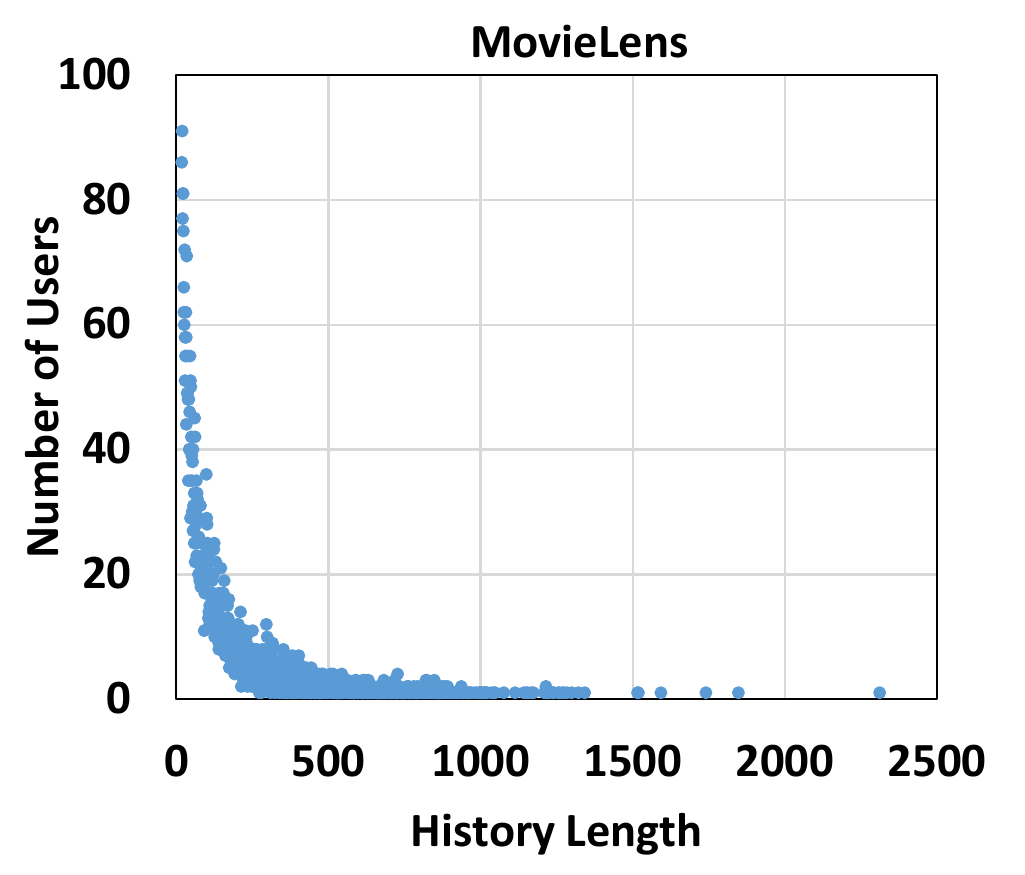}
		\vspace{-15pt}
	\end{subfigure} 
	\begin{subfigure}[b]{0.23\textwidth}
		\centering
		\includegraphics[width=\textwidth]{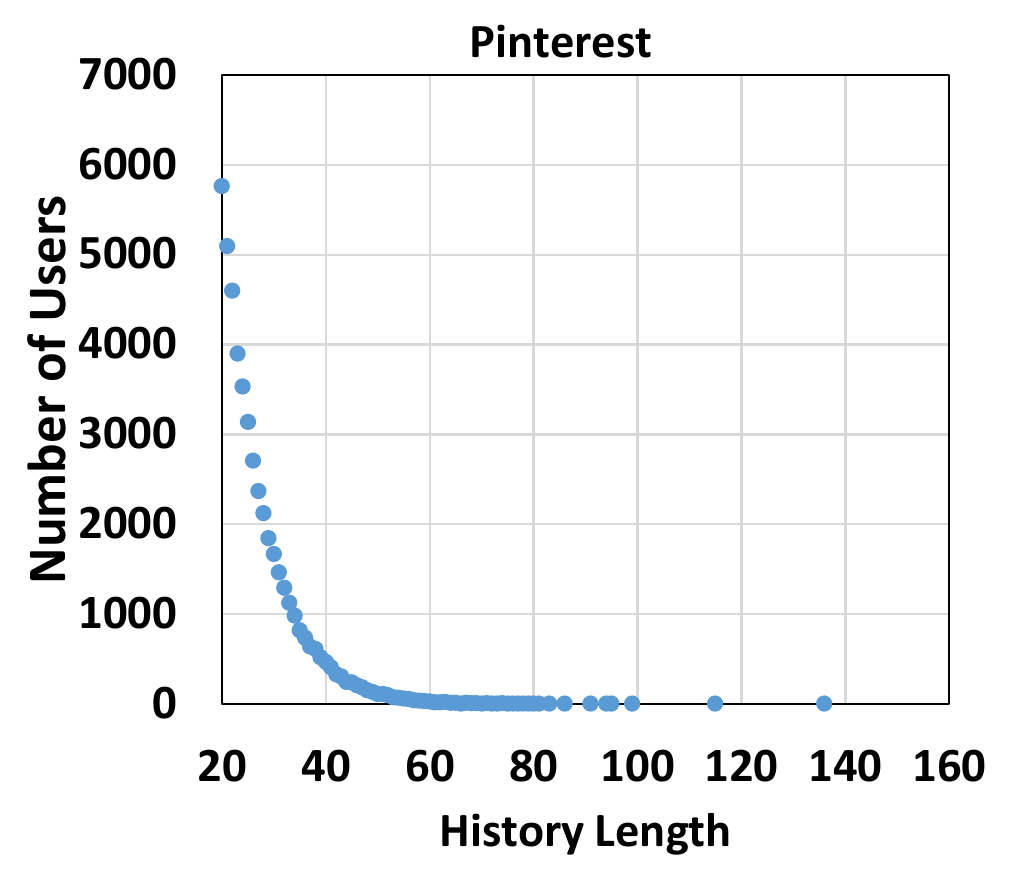}
		\vspace{-15pt}
	\end{subfigure} 
	\caption{The distribution of user history length on our experimented MovieLens and Pinterest datasets.}
	\vspace{-15pt}
	\label{fig:history_dist}
\end{figure}

To justify this point, we show the distribution of user history length on our experimented MovieLens and Pinterest datasets in Figure~\ref{fig:history_dist}. 
We can see that for both real-world datasets, the history length of users varies a lot; specifically, the $(mean, variance)$ of user history length are $(166,37145)$ and $(27,57)$ for MovieLens and Pinterest, respectively. 
Taking the left subfigure of MovieLens data as an example, the average length for all users is 166, while the maximum length is 2313. Which means, the average attention weight of the most active user is 1/2313, about 14 times fewer than that of average users (\ie 1/166). Such a large variance on attention weights will cause problems in optimizing the item embeddings of the model. \vspace{+5pt}

\noindent\textbf{The NAIS Model}. We now present our final design for the NAIS model. As analyzed above, the weak performance of Design 3 comes from the softmax, which performs $L_1$ normalization on attention weights and results in large variance on attention weights of different users. To address the problem, we propose to smooth the denominator of softmax, so as to lessen the punishment on attention weights of active users and meanwhile decrease the variance of attention weights. Formally, the predictive model of NAIS is as follows:
\begin{equation} \label{eq:NAIS}
	\begin{aligned}
		\hat{y}_{ui} &= \textbf{p}_i^T ( \sum_{j\in \mathcal{R}^+_u \setminus \{i\}} a_{ij} \textbf{q}_j), \\
		a_{ij} &= \frac{\exp{(f(\textbf{p}_i, \textbf{q}_j))}}{[\sum_{j\in \mathcal{R}^+_u\setminus \{i\}} \exp{(f(\textbf{p}_i, \textbf{q}_j))]^\beta }},
	\end{aligned}
\end{equation}
where $\beta$ is the smoothing exponent, a hyperparameter to be set in the range of $[0,1]$. Obviously, when $\beta$ is set to 1, it recovers the softmax function; when $\beta$ is smaller than 1, the value of denominator will be suppressed, as a result the attention weights will not be overly punished for active users. 
Although the probabilistic explanation of attention network is broken with $\beta<1$, we empirically find that it leads to a performance much better than using the standard softmax (see Section \ref{ss:hyper-parameter} for experiment results). 
We use the term ``NAIS-concat'' and ``NAIS-prod'' to denote the NAIS model that uses $f_{concat}$ and $f_{prod}$ as the attention function, respectively (\cf Equation~(\ref{eq:f})). 

Moreover, our NAIS model can be viewed under the recently proposed \textit{Neural Collaborative Filtering} (NCF) framework~\cite{NCF}, as illustrated in Figure~\ref{fig:NAIS}. Differing from the user-based NCF models that use one-hot user ID as the input feature, our NAIS model uses multi-hot interacted items as the input feature for a user. Together with the carefully designed attention network as the hidden layer, our NAIS model can be more intuitively understood as performing item-to-item CF. 

\subsection{Optimization}
To learn a recommender model, we need to specify an objective function to optimize. As we deal with implicit feedback where each entry is a binary value 1 or 0, we can deem the learning of a recommender model as a binary classification task. Similar to the previous work on Neural CF \cite{NCF}, we treat the observed user-item interactions as positive instances, sampling negative instances from the remaining unobserved interactions. Let $\mathcal{R}^+$ and $\mathcal{R}^-$ denote the set of positive and negative instances, respectively,  we minimize the regularized log loss defined as follows:
\begin{equation} \small
	L = -\frac{1}{N} \big(\sum_{(u,i)\in \mathcal{R}^+} \log \sigma(\hat{y}_{ui}) + \sum_{(u,i)\in \mathcal{R}^-} \log (1- \sigma(\hat{y}_{ui}))\big)
	+ \lambda ||\Theta||^2
\end{equation}
where $N$ denotes the number of total training instances, and $\sigma$ is a sigmoid function that converts a prediction $\hat{y}_{ui}$ to a probability value denoting the likelihood that $u$ will interact with $i$. The hyper-parameter $\lambda$ controls the strength of $L_2$ regularization to prevent overfitting, and $\Theta=\{\{\textbf{p}_i\}, \{\textbf{q}_i\}, \textbf{W}, \textbf{b}, \textbf{h}\}$ denotes all trainable parameters. We are aware of other options of objective functions, such as the pointwise regression~\cite{fastMF,wang2017learning} and pairwise ranking~\cite{BPR,zhaoIJCAI17} losses, can also be employed to learn NAIS for implicit feedback. As the focus of the work is to show the effectiveness of NAIS, especially on the improvement over FISM to justify the usage of attention, we leave the exploration of other objective functions as future work. 

To optimize the objective function, we adopt Adagrad~\cite{Adagrad}, a variant of \textit{Stochastic Gradient Descent} (SGD) that applies an adaptive learning rate for each parameter. It draws a stochastic sample from all training instances, updating the related parameters towards the negative direction of their gradients. 
We use the mini-batch version of Adagrad to speedup the training process, and the generation of a mini-batch is detailed in Section~\ref{ss:settings} of experimental settings. 
In each training epoch, we first generate all negative instances, and then feed them together with positive instances into the training algorithm for parameter updates.
This leads to much faster training than sampling the negative instance on-the-fly (as done in Bayesian Personalized Ranking~\cite{BPR}) when training on GPU platforms, since it avoids the unnecessary switch between GPU (for parameter updating) and CPU (for negative sampling).
Specifically, for each positive instance $(u,i)$, we randomly sample $X$ items that $u$ has never interacted before as negative instances. In our experiments we set $X$ as 4, an empirical number that has shown good performance for neural CF methods~\cite{NCF}. \vspace{+5pt}

\textbf{Pre-training}. Due to the non-linearity of neural network model and non-convexity of the objective function (\wrt all parameters), optimization using SGD can be easily trapped to local minimums of poor performance. As such, the initialization of model parameters plays a vital role in the model's final performance.
Empirically, when we try to train NAIS from random initialization, we find it converges slowly and leads to a final performance slightly better than FISM. 
We hypothesize that it is due to the difficulty of optimizing the attention network and item embeddings simultaneously. Since the outputs of attention network rescale item embeddings, jointly training them may result in the co-adaption effect, which slows down the convergence. For example, a training epoch may decrease an attention weight $a_{ij}$ but increase the embedding product $\textbf{p}_i^T\textbf{q}_j$, resulting in only a small progress in updating the prediction score. 

To address the practical issue in training NAIS, we pre-train NAIS with FISM, using the item embeddings learned by FISM to initialize that of NAIS. 
Since FISM does not have the co-adaption issue, it can learn item embeddings well in encoding item similarity. 
As such, using FISM embeddings to initialize NAIS can greatly facilitate the learning of the attention network, leading to faster convergence and better performance. 
With such a meaningful initialization of item embeddings, we can simply initialize the attention network with a random Gaussian distribution. 

\subsection{Discussions}
\label{ss:discussions}
In this subsection, we discuss three properties of NAIS, namely, its time complexity, ease to support online personalization, and the two options for attention function. 

\textbf{Time Complexity Analysis.} We analyze the time complexity of the predictive model of NAIS, \ie Equation (\ref{eq:NAIS}). This directly reflects the time cost of NAIS in testing (or recommendation), and the time cost of training should be proportional to that of testing. 
The time complexity of evaluating a prediction $\hat{y}_{ui}$ with FISM (\cf Equation (\ref{eq:FISM})) is $O(k|\mathcal{R}_u^+|)$, where $k$ denotes the embedding size and $|\mathcal{R}_u^+|$ denotes the number of historical interactions of user $u$. Compared to FISM, the additional cost of evaluating a prediction with NAIS comes from the attention network. Let $a$ denote the attention factor, then we can express the time complexity of evaluating $f(\textbf{p}_i, \textbf{q}_j)$ as $O(ak)$. Since the denominator of softmax (and our proposed smoothed variant of softmax) needs to traverse over all items in $\mathcal{R}_u^+$, the time complexity of evaluating an $a_{ij}$ is $O(ak|\mathcal{R}_u^+|)$. As such, a direct implementation of NAIS model takes time $O(ak|\mathcal{R}_u^+|^2)$, since we need to evaluate $a_{ij}$ for each $j$ in $|\mathcal{R}_u^+|$. However, considering the denominator term is shared across the computation of all items in $\mathcal{R}_u^+$, we only need to compute it once and cache it for all evaluations of $a_{ij}$ (where j is in  $\mathcal{R}_u^+$). As such, the overall time complexity of evaluating a NAIS prediction can be reduced to $O(ak|\mathcal{R}_u^+|)$, which is $a$ times of that of FISM. \vspace{+5pt}

\textbf{Support for Online Personalization.} The offline training of a recommender model provides personalized recommendation based on a user's past history. For online personalization, we consider the practical scenario that a user has new interactions streaming in, and the recommender model needs to refresh the top-K recommendation for the user instantaneously~\cite{fastMF,zhaoTKDE2016}. Since it is prohibitive to perform model re-training in real-time\footnote{To the authors' knowledge, the current industrial servings of recommender systems usually perform model re-training on a daily basis. }, an alternative solution is to perform local updates on model parameters based on the new feedback only. This is the common strategy used by user-based CF model, such as matrix factorization~\cite{fastMF}. However, we argue that even local updates on parameters are difficult to achieve in practice. The key difficulty is that users may have concurrent interactions on an item. As such, separately performing local updates on a per interaction basis will result in collision, and it is non-trivial to resolve the collision in a distributed setting in real-time. 

Instead of updating model parameters to adapt new interactions, NAIS can refresh the representation vector of a user without updating any model parameter, reducing the difficulty of provide online personalization services. This is attributed to the item-based CF mechanism that characterizes a user with her interaction history, rather than her ID. Specifically in NAIS, a user's representation vector is aggregated by a weighted sum on item embeddings, which allows a nice decomposable evaluation on a prediction. For example, let's assume user $u$ has a new interaction on item $t$. To refresh the prediction of $u$ on a candidate item $i$ (\ie $\hat{y}_{ui}$), instead of computing $\hat{y}_{ui}$ from scratch (\ie following Equation~(\ref{eq:NAIS})), we only need to evaluate the score of $a_{it} \textbf{p}_i^T \textbf{q}_t$, and then sum it with the old prediction of $\hat{y}_{ui}$. With the cache of the denominator of softmax, the refresh of $\hat{y}_{ui}$ can be done in $O(ak)$ time. This is much more efficient than performing local updates with MF~\cite{fastMF} (for which the time complexity is $O(k^2+|\mathcal{R}_u^+|k)$), since $a$ is usually a small number (typically set to be the same as $k$). \vspace{+5pt}

\textbf{Options for Attention Function.}
The two choices of attention function differ in the construction of input: the first choice $f_{concat}$ simply concatenates $\textbf{p}_i$ and $\textbf{q}_j$ to learn the attention weight $w_{ij}$~\cite{bahdanau2014neural}, while second choice $f_{prod}$ feeds the element-wise product of $\textbf{p}_i$ and $\textbf{q}_j$ into the attention network~\cite{ijcai2017-afm}. Analytically speaking, since the attention weight $w_{ij}$ is to score the interaction $\textbf{p}_i^T \textbf{q}_j$, using the element-wise product $\textbf{p}_i\odot \textbf{q}_j$ as input may facilitate the hidden layer in learning the attention function (since $\textbf{p}_i^T \textbf{q}_j=\textbf{1}^T(\textbf{p}_i\odot \textbf{q}_j)$); as a downside, it may also cause some information loss unintentionally, since the original information encoded in $\textbf{p}_i$ and  $\textbf{q}_j$ are discarded. In contrast, $f_{concat}$ leverages the original information encoded in $\textbf{p}_i$ and $\textbf{q}_j$ to learn their interaction weight, which has no information loss; however, due to the numerical gap between the concatenation $[\textbf{p}_i,\textbf{q}_j]^T$ and element-wise product $\textbf{p}_i\odot\textbf{q}_j$, it may lead to slower convergence. We will empirically compare the two choices of attention function in the experiments section.

\section{Experiments}
\label{sec:exper}

In this section, we conduct experiments with the aim of answering the following  research questions: \vspace{+5pt}

\noindent\textbf{RQ1} Are our proposed attention networks useful for providing more accurate recommendations?

\noindent\textbf{RQ2} How do our proposed NAIS methods perform compared with state-of-the-art recommendation methods?

\noindent\textbf{RQ3} What are the key hyper-parameters for NAIS and how do they impact NAIS's performance? \vspace{+5pt}

\noindent In what follows, we first present the experimental settings, followed by results answering the above questions. 

\subsection{Experimental Settings}
\label{ss:settings}
\noindent \textbf{Datasets and Evaluation Protocol.} We adopt the same MovieLens and Pinterest datasets as the ones used in the NCF paper \cite{NCF}.
Since both datasets have some pre-processing steps such as removing sparse users and train-test splitting, we directly evaluate on the processed data\footnote{The processed datasets are directly downloaded from: https://github.com/hexiangnan/neural\_collaborative\_filtering}. Table~\ref{tab:dataset} summarizes the statistics of the two datasets. 
More details on the generation of the two datasets have been elaborated in \cite{NCF}, so we do not restate them. 
Note that during training each interaction is paired with $4$ negative instances, thus the number of training instances is much more than the number of interactions. 

We adopt the leave-one-out evaluation protocol~\cite{BPR,NCF}, which holds out the latest interaction of each user as the testing data and uses the remaining interactions for training. Specifically, each testing instance is paired with $99$ randomly sampled negative instances; then each method outputs prediction scores for the $100$ instances ($1$ positive plus $99$ negatives), and the performance is judged by \textit{Hit Ratio} (HR)~\cite{deshpande2004item} and \textit{Normalized Discounted Cumulative Gain} (NDCG)~\cite{TriRank} at the position 10. 
Both metrics have been widely used to evaluate top-K recommendation~\cite{FISM} and ranking systems~\cite{he2017birank} in information retrieval literatures. 
We report the average scores for all users, where $HR@10$ can be interpreted as a recall-based measure that indicates the percentage of users are successfully recommended (\ie the positive instance appears in top-10), and $NDCG@10$ is a precision-based measure that accounts for the predicted position of the positive instance, the larger the better. \vspace{+5pt}

\begin{table}[t]
	\begin{center}
		\caption{\textbf{Statistics of the evaluation datasets.}}\vspace{-10pt}
		\small
		\label{tab:dataset}
		\begin{tabular}{ | l | c | c | c | c | }
			\hline
			\textbf{Dataset} & \textbf{Interaction\#} & \textbf{Train\#} & \textbf{Item\#} & \textbf{User\#}  \\ \hline
			MovieLens	& 1,000,209 & 4,970,845 & 3,706 & 6,040  \\ \hline
			Pinterest	& 1,500,809	& 7,228,110 & 9,916 & 55,187 \\ \hline
		\end{tabular}
		\vspace{-15pt}
	\end{center}
\end{table}

\begin{figure*}[t]
	\centering
	\begin{subfigure}[b]{0.24\textwidth}
		\centering
		\includegraphics[width=\textwidth]{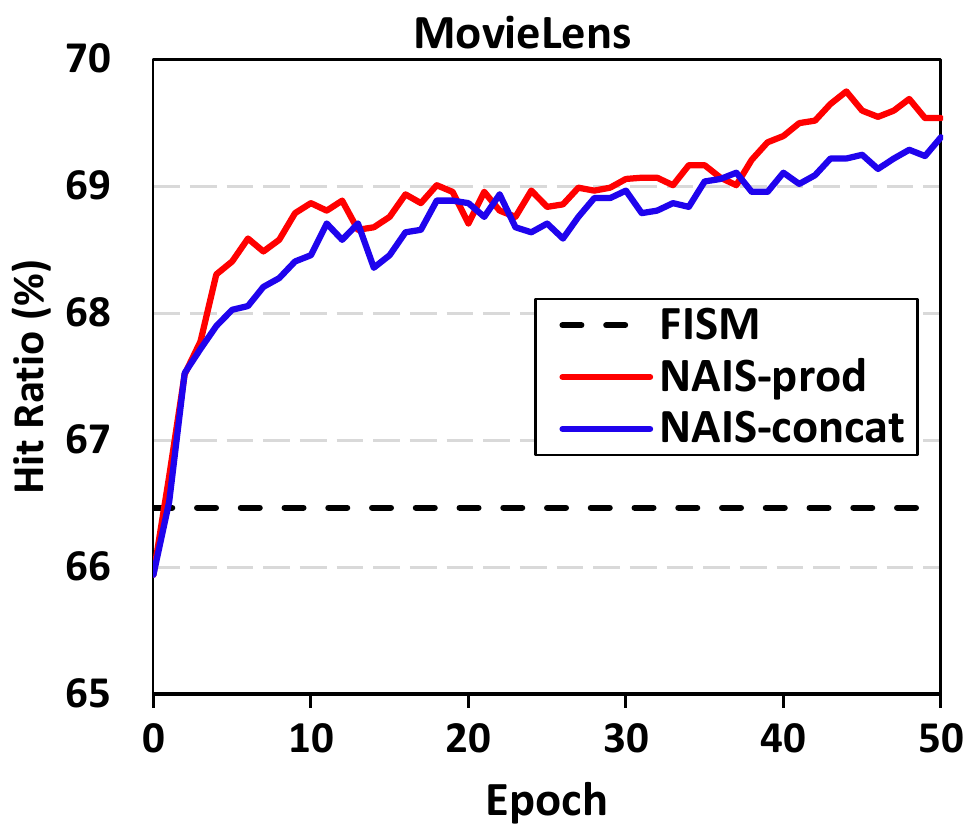}
		\vspace{-18pt}
		\caption{MovieLens --- HR}
		\label{fig:attention-ml-hr}
	\end{subfigure} 
	\begin{subfigure}[b]{0.24\textwidth}
		\centering
		\includegraphics[width=\textwidth]{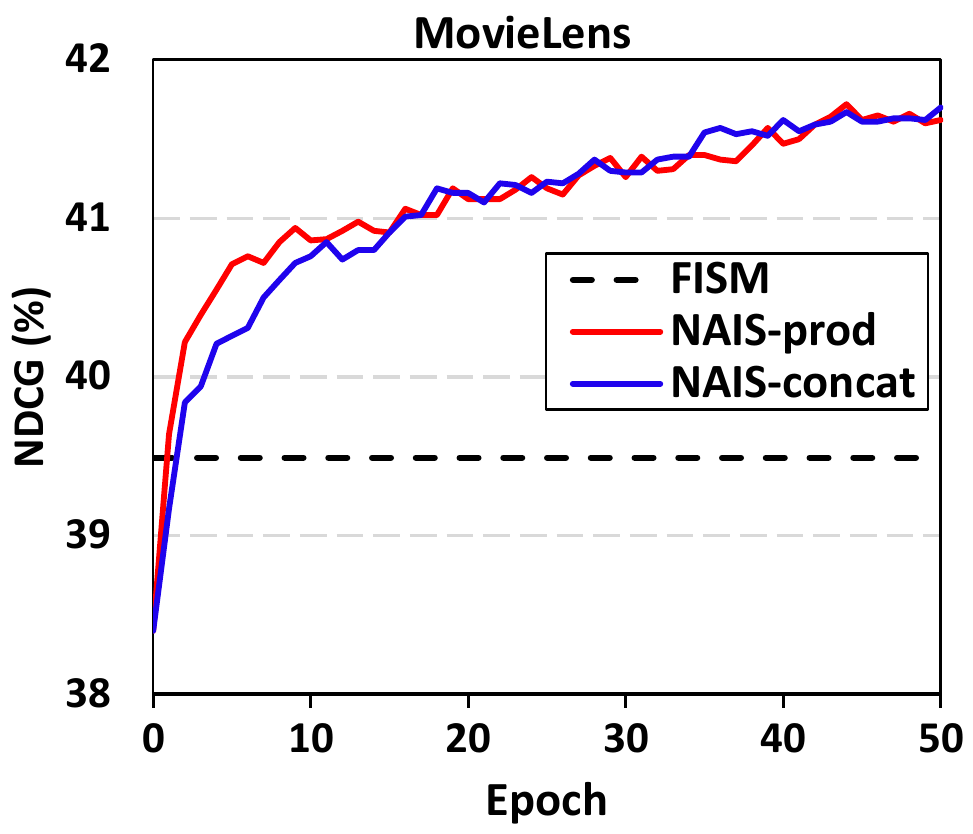}
		\vspace{-18pt}
		\caption{MovieLens --- NDCG}
		\label{fig:attention-ml-ndcg}
	\end{subfigure} 
	\begin{subfigure}[b]{0.24\textwidth}
		\centering
		\includegraphics[width=\textwidth]{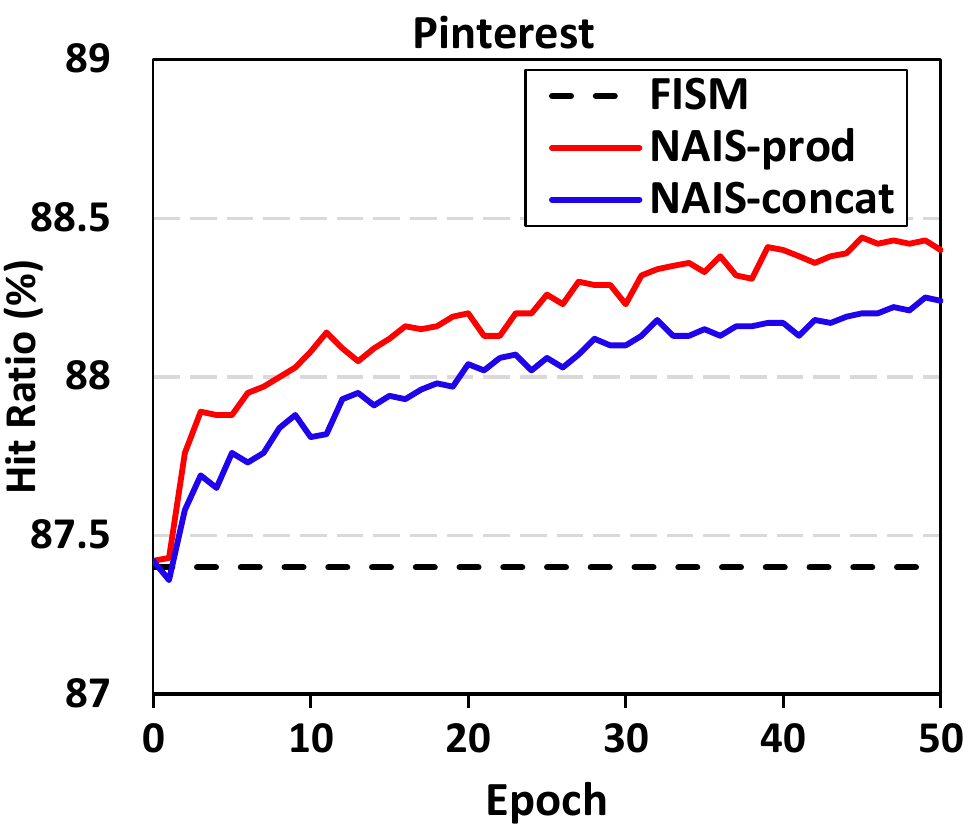}
		\vspace{-18pt}
		\caption{Pinterest --- HR}
		\label{fig:attention-pinterest-hr}
	\end{subfigure} 
	\begin{subfigure}[b]{0.24\textwidth}
		\centering
		\includegraphics[width=\textwidth]{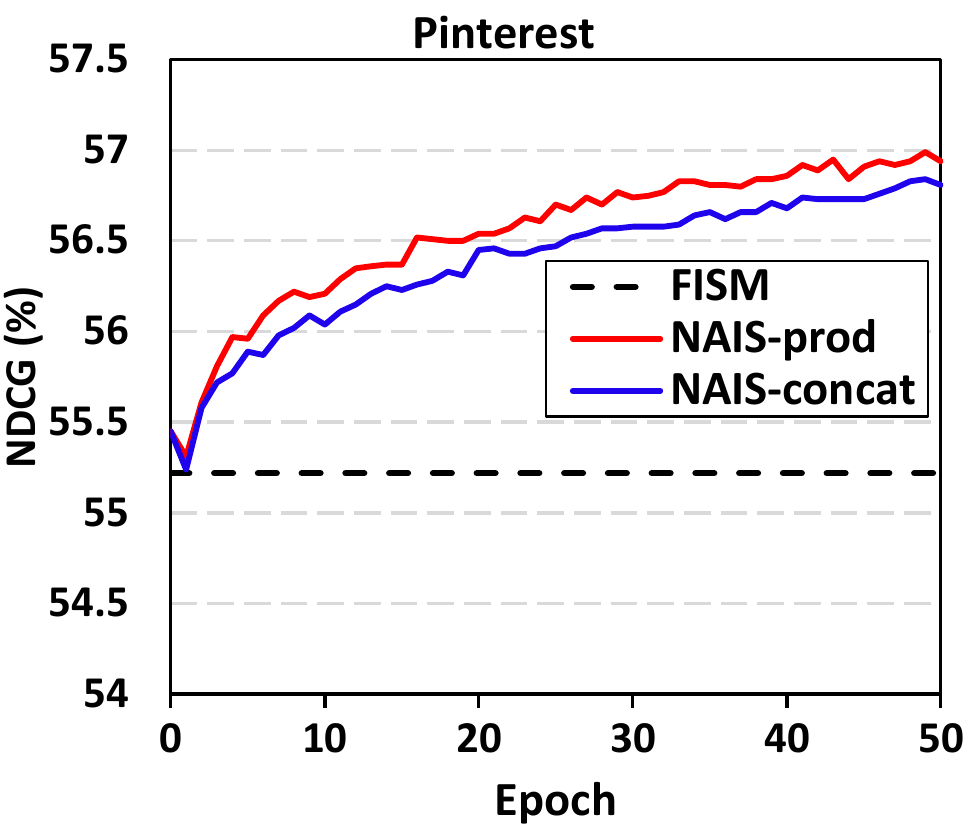}
		\vspace{-18pt}
		\caption{Pinterest --- NDCG}
		\label{fig:attention-pinterest-ndcg}
	\end{subfigure} 
	\caption{Testing performance of FISM, NAIS-prod, and NAIS-concat of embedding size 16 in each epoch.}
	\vspace{-10pt}
	\label{fig:epoch}
\end{figure*}

\noindent \textbf{Baselines.} We compare NAIS with the following item recommendation methods:

\textbf{Pop}. This is a non-personalized method to benchmark the performance of the top-K recommendation task. It ranks items by their popularity, judged by the number of interactions that an item received. 

\textbf{ItemKNN}~\cite{ICF}. This is the standard item-based CF method as formulated in Equation (\ref{eq:itemCF}). We use consine similarity to measure $s_{ij}$. We experiment with different numbers of nearest item neighbors to consider, finding using all neighbors lead to best results. 

\textbf{FISM}~\cite{FISM}. This is a state-of-the-art item-based CF model as formulated in Equation (\ref{eq:FISM}). We test $\alpha$ from $0$ to $1$ with a step size of $0.1$, finding a value of 0 leads to best result on both datasets (the variance is actually small when $\alpha$ is smaller than 0.6).  

\textbf{MF-BPR}~\cite{BPR}. MF-BPR learns MF by optimizing the pairwise \textit{Bayesian Personalized Ranking} (BPR) loss. This method is a popular choice for building a CF recommender from implicit feedback. 

\textbf{MF-eALS}~\cite{fastMF}. This method also learns a MF model, but optimizes a different pointwise regression loss that treats all missing data as negative feedback with a smaller weight. The optimization is done by an \textit{element-wise Alternating Learning Square} (eALS) algorithm. 

\textbf{MLP}~\cite{NCF}. This method applies a multi-layer perceptron (MLP) above user and item embeddings to learn the scoring function from data. We employ a 3-layer MLP and optimize the same pointwise log loss, which was reported to perform well on the two datasets. 

We have deliberately chosen the above methods to cover a diverse range of recommendation methods: ItemKNN and FISM are representative of item-based CF approaches to validate the utility of our attention-argument modeling,  MF-BPR and MF-eALS are competitive user-based CF approaches to evidence the state-of-the-art performance of recommendation from implicit feedback, and MLP is a recently proposed deep neural network-based CF method. 
Note that we focus on the comparison of single CF models. As such, we do not further compare with NeuMF which achieves the best performance in the NCF paper, since NeuMF is an ensemble method that fuses MF and MLP in the latent space. \vspace{+5pt} 

\noindent \textbf{Parameter Settings.} 
For each method, we first train it without regularization; if overfitting is observed (\ie training loss keeps decreasing but the performance becomes worse), we then tune the regularization coefficient $\lambda$ in the range of $[10^{-6}, 10^{-5}..., 1]$. 
The validation set is consisted of a randomly drew interaction for each user.  
For the embedding size $k$, we test the values of $[8, 16, 32, 64]$, and set the attention factor $a$ same as the embedding size in each setting. 
For a fair comparison with FISM, we optimize it with the same pointwise log loss using the same Adagrad learner. We find that using the item embeddings learned by FISM to initialize NAIS (\ie the pre-training step) leads to slightly better performance but much faster convergence. Without special mention in texts, we report the performance of NAIS with following default settings: 1) $\beta=0.5$, 2) $k=a=16$, 3) $\lambda=0$, 4) Adagrad with a learning rate of 0.01, and 5) pre-training with FISM embeddings. \vspace{+5pt}

\noindent \textbf{Implementation Details.} We implement NAIS using TensorFlow\footnote{Our implementation codes are available at \url{https://github.com/AaronHeee/Neural-Attentive-Item-Similarity-Model}}. Since in the input layer an item (user) is represented as a one-hot (multi-hot) vector where most entries are zeros, for efficiency and memory concern, we adopt sparse representation that stores the IDs of non-zero entries only. 
Here an implementation challenge is that different users have different number of non-zero entries, while TensorFlow requires all training instances of a batch must be of the same length (same as other programming tools for deep learning like Theano). To tackle the challenge, a widely adopted solution is to use the \textit{masking} trick, which adds masks (\ie pseudo non-zero entries) to ensure all instances of a batch have a same length (\ie the maximum length of instances of the batch). However, we find this solution is very time-consuming on CF datasets, as some active users may have interacted with over thousands of items, making a sampled mini-batch very large. To address the issue, we innovatively form a mini-batch as all training instances of a randomly sampled user, rather than randomly sampling a fixed number of training instances as a mini-batch. This trick of user-based mini-batch has two advantages: 1) no mask is used thus it is much faster (empirically 3X speedup over the masking trick), and 2) no batch size needs to be specified which refrains the pain of tuning the batch size.
Moreover, the recommendation performance remains the same according to our experiments. \vspace{+5pt}

\begin{table}[h]
	\begin{center}
		\caption{\textbf{Training time per epoch (seconds) of methods that are implemented using TensorFlow.}}\vspace{-10pt}
		\label{tab:time}
		\begin{tabular}{ | l | c | c | c | c | }
			\hline
			\textbf{Methods} & \textbf{MovieLens} & \textbf{Pinterest}  \\ \hline
			\textbf{MF-BPR}	& 24.4 s & 17.3 s   \\ \hline
			\textbf{MLP}	& 125.8 s	& 155.8 s  \\ \hline
			\textbf{FISM}	& 238.3	s & 353.3 s  \\ \hline
			\textbf{NAIS\_concat}	& 455.2 s	& 525.6 s  \\ \hline
			\textbf{NAIS\_prod}	& 428.5	s &  485.2 s  \\ \hline
		\end{tabular}
	\end{center}
\end{table}

\noindent\textbf{Training Time}. Table \ref{tab:time} shows the training time per epoch of NAIS and baselines that are implemented with TensorFlow. A training epoch is defined as training $5|\mathcal{R}^+|$ instances, since the negative sampling ratio is 4. The running environment is a server with Intel Xeon CPU E5-2630 @ 2.20GHz and 64GB memory. 
Note that the running time of ItemKNN and MF-eALS are not shown since they are implemented with Java, which are not comparable with other methods. We can see that item-based CF methods (FISM and NAIS) take longer training time than user-based CF methods (MF-BPR and MLP). This is reasonable, since user-based methods use an ID only to represent a user in the input layer, while item-based methods use  interacted items to represent a user. MLP uses more time than MF-BPR, since it has three more hidden layers than MF-BPR. 
Moreover, the two NAIS methods take longer time than FISM, due to the additional use of attention network. The additional time cost is quite acceptable, which is roughly 0.8 times of the training time of FISM. Among the two NAIS methods, NAIS\_concat takes slightly longer time than NAIS\_prod, since concatenation increases the input dimension while product does not. 

\subsection{Effectiveness of Attention Networks (RQ1)}
Technically speaking, our NAIS model enhances FISM by replacing the constant weight (\ie $1/|R_u^+|^\alpha$) of an estimated item-item similarity (\ie $\textbf{p}_i^T \textbf{q}_j$) with a variable weight learned by an attention network. To demonstrate the efficacy of our designed attention networks, we first run FISM until convergence, and then use FISM embeddings to initialize NAIS for training the attention network. 

Figure \ref{fig:epoch} shows the stable performance of FISM and the scores of our two NAIS methods at embedding size 16 in each epoch. We can clearly see the effectiveness of using attention networks. 
Specifically, the initialized performance of NAIS are close to FISM, while by training the attention network, the two NAIS methods improve over FISM significantly. 
Here we show the performance of 50 epochs only, and further training on NAIS can lead to even better performance. Upon convergence (results can be found in Table \ref{tab:overall_performance}), both NAIS methods achieve a relative improvement of $6.3\%$ and $3.6\%$ over FISM in terms of NDCG on MovieLens and Pinterest, respectively. We believe the improvements on recommendation accuracy stem from the strong representation power of NAIS.
Moreover, we find that NAIS-prod converges faster than NAIS-concat (while their final performance are close). This confirms our analysis in Section~\ref{ss:discussions} by providing empirical evidence that feeding $\textbf{p}_i\odot\textbf{q}_j$ into the attention network can facilitate learning the weight of $\textbf{p}_i^T\textbf{q}_j$.

\subsubsection{Qualitative Analysis}

\begin{figure*}[t]
	\centering
	\begin{subfigure}[b]{0.245\textwidth}
		\centering
		\includegraphics[width=\textwidth]{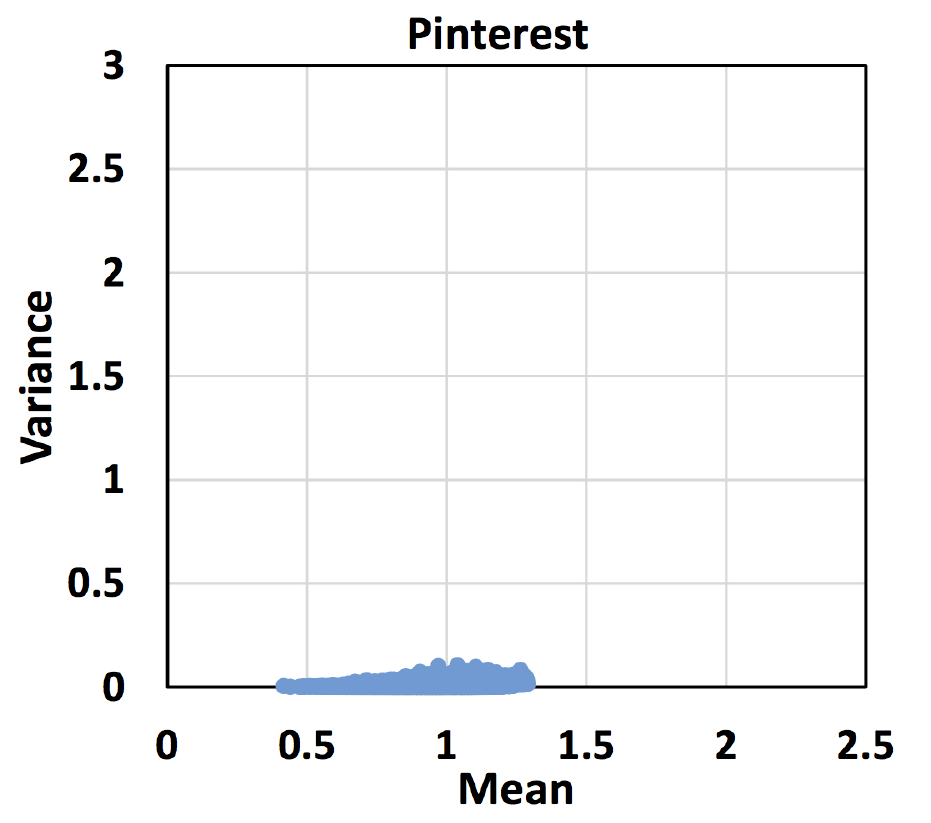}
		\vspace{-10pt}
		\caption{Epoch 1}
		\label{fig:attention-dist-pinterest1}
	\end{subfigure} 
	\begin{subfigure}[b]{0.245\textwidth}
		\centering
		\includegraphics[width=\textwidth]{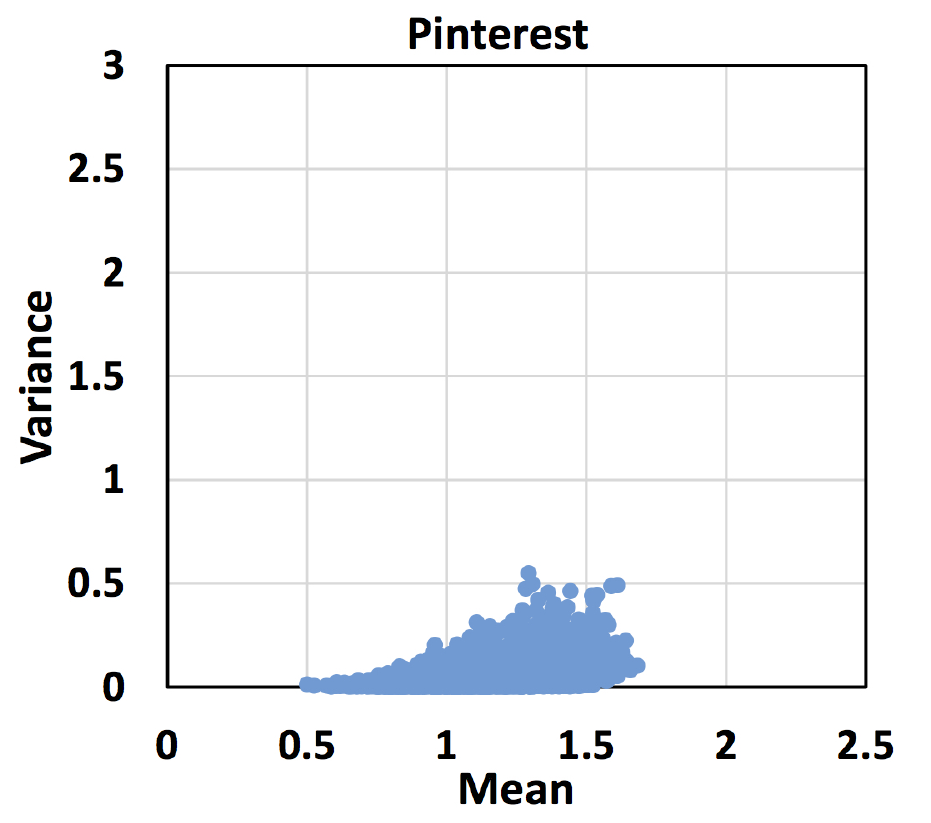}
		\vspace{-10pt}
		\caption{Epoch 10}
		\label{fig:attention-dist-pinterest2}
	\end{subfigure} 
	\begin{subfigure}[b]{0.245\textwidth}
		\centering
		\includegraphics[width=\textwidth]{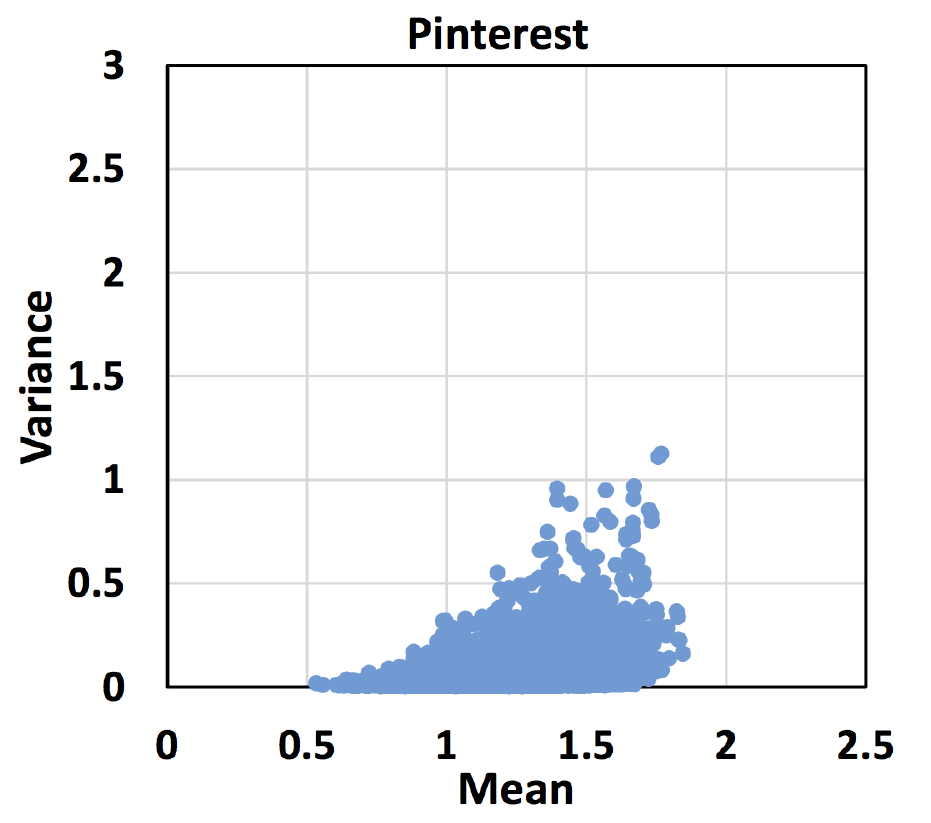}
		\vspace{-10pt}
		\caption{Epoch 20}
		\label{fig:attention-dist-pinterest3}
	\end{subfigure} 
	\begin{subfigure}[b]{0.245\textwidth}
		\centering
		\includegraphics[width=\textwidth]{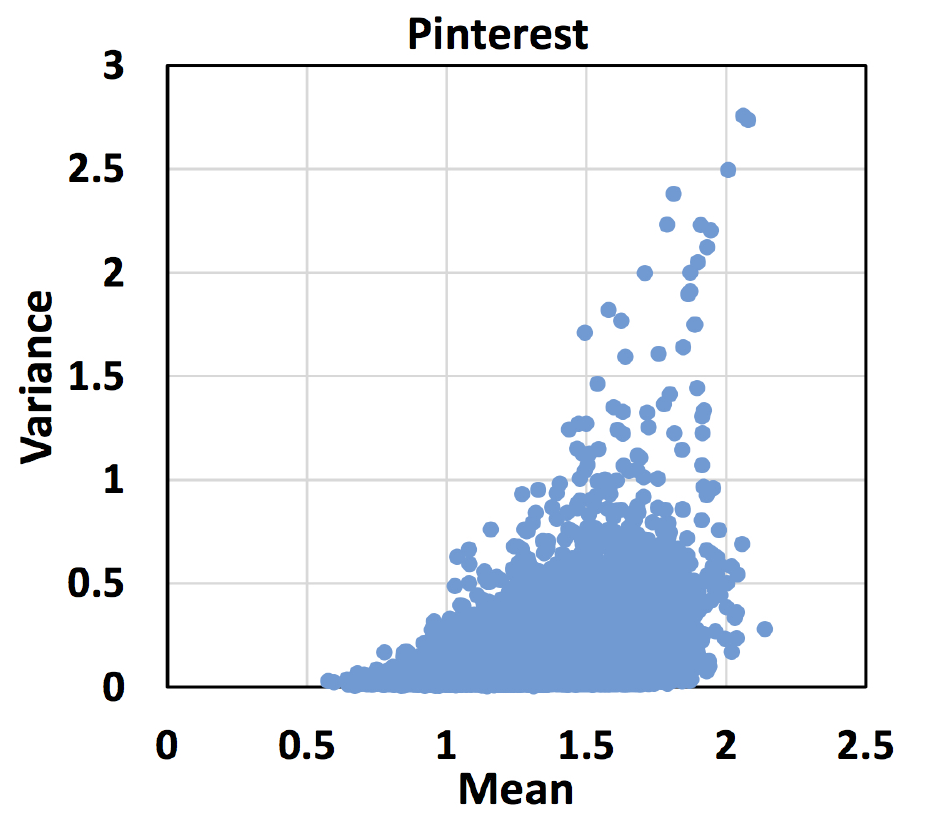}
		\vspace{-10pt}
		\caption{Epoch 40}
		\label{fig:attention-dist-pinterest4}
	\end{subfigure} 
	\caption{The scatter plot of mean (x-axis) and variance (y-axis) of attention weights learned by NAIS-prod at different epochs. Each scatter point denotes the prediction of a testing point in Pinterest.}
	\vspace{-10pt}
	\label{fig:attention-dist-pinterest}
\end{figure*}

Here we provide some qualitative analysis on the attention weights to show their learnability and interpretability. 

First, it is interesting to see how do the attention weights evolve during training. 
However, a prediction of $\hat{y}_{ui}$ has $|\mathcal{R}_u^+|$ attention weights, and it is difficult to plot the attention weights for all predictions. 
Instead, we record the statistics --- mean and variance --- of the attention weights of a prediction, and effective learning of the attention network is evidenced by a large variance (note that the variances of FISM are 0). 
Figure \ref{fig:attention-dist-pinterest} shows the scatter plot of the statistics learned by NAIS-prod at different epochs in Pinterest, where each scatter point denotes the prediction of a testing instance. We can see that in the initial phase of training (Epoch 1), the points are concentrated near x-axis, \ie variances are close to zero. With more training epochs, the points become more dispersive along the y-axis, and many points start to get a high variance. 
Together with Figure \ref{fig:epoch} which shows more training epochs lead to better performance, we can conclude that the attention weights have been properly trained to be more distinguishable for historical items. 
This reveals the reason of NAIS improving over FISM, justifying our key argument of this work that the historical items of a user do not contribute equally in a prediction. 

\begin{table}[h]
	\begin{center}
		\caption{\textbf{Attention weights breakdown of a sampled user on target item \#1382 in Pinterest. The user has four historical items which are shown in column 1 to 4, and the last column denotes the prediction score (after sigmoid).}} \vspace{-5pt}
		\label{tab:case}
		\begin{tabular}{| l | c | c | c | c | c |} \hline
			\textbf{Item ID} & \textbf{\#131} & \textbf{\#894} & \textbf{\#1534} & \textbf{\#3157} & $\sigma (\hat{y}_{ui})$ \\ \hline
			\textbf{FISM} 	& 0.25 & 0.25 & 0.25 & 0.25 & 0.17 \\ \hline
			\textbf{NAIS-prod} & 0.03 & 0.52 & 0.22	& 0.23 & 0.81  \\ \hline
		\end{tabular}
	\end{center}
\end{table}

Second, we show a case study on the attention weights of a prediction of a sampled user in Table~\ref{tab:case}. The weights have been $L_1$ normalized to make a clear comparison with FISM, which assumes a uniform weight on the historical items. 
In this example, the target item $\#1382$ is a positive example in the testing set and should be scored larger. We can see that FISM weights all historical items (more precisely, their interactions with the target item) uniformly, which leads to a relatively smaller prediction score. In contrast, NAIS-prod assigns a higher weight on item $\#894$ and a lower weight on item $\#131$, successfully scoring the target item $\#1382$ larger, which is desired. To demonstrate the rationality, we further investigate the content of these items (\ie Pinterest images). We find that both the target item $\#1382$ and the highest attended item $\#894$ are about natural scenery, while the lowest attended item $\#131$ is a family photo. This is as expected, because when predicting a user's preference on a target item, her historical items of the same category should have a larger impact than other less relevant items. 
This well justifies our motivating example in introduction, providing evidence on the correlation of the attention weights and the characteristics of items.

\subsubsection{Effect of Pre-training}
\begin{table}[h]
	\begin{center}
		\caption{\textbf{Performance of NAIS methods with (w/) and without (w/o) FISM pre-training at embedding size 16.}} \vspace{-5pt}
		\label{tab:pretrain}
		\begin{tabular}{| l | c | c | c | c |} \hline
			& \multicolumn{2}{c|}{\textbf{MovieLens}} & \multicolumn{2}{c|}{\textbf{Pinterest}} \\ \hline 
			\textbf{Methods} & \textbf{HR} & \textbf{NDCG} & \textbf{HR} & \textbf{NDCG} \\ \hline
			\textbf{FISM} 	& 66.47 & 39.49 & 87.40 & 55.22 \\ \hline\hline
			\textbf{NAIS-concat w/o pre-training} & 67.77 & 40.41 & 87.90	& 56.23  \\ \hline
			\textbf{NAIS-concat w/ pre-training} & \textbf{69.72} & \textbf{41.96} & \textbf{88.44}	& \textbf{57.20}  \\ \hline\hline
			\textbf{NAIS-prod w/o pre-training} & 68.04 & 40.55 &	87.90 & 56.04 \\ \hline
			\textbf{NAIS-prod w/ pre-training} & \textbf{69.69} & \textbf{41.94} &	\textbf{88.44} & \textbf{57.22} \\ \hline
		\end{tabular}
	\end{center}
\end{table}

To demonstrate the effect of pre-training (\ie, using the embeddings learned by FISM as model initialization), we show the performance of NAIS with and without pre-training at embedding size 16 in Table~\ref{tab:pretrain}. 
Note that the hyper-parameters of NAIS without pre-training have been separately tuned. 
As can be seen, by pre-training the two NAIS methods with FISM embeddings, both methods are improved significantly. Besides the performance improvements, NAIS methods with pre-training have a faster convergence rate than random initialization. 
This points to the positive effect of using FISM embeddings to initialize NAIS. Moreover, training NAIS from scratch leads to a performance better than FISM, which further verifies the usefulness of the attention network. 

\subsection{Performance Comparison (RQ2)}
\label{ss:performance}
We now compare the performance of NAIS with other item recommendation methods. 
For these embedding-based methods (MF, MLP, FISM, and NAIS), the embedding size controls their modeling capability; as such, we set it to 16 for all methods for a fair comparison. In later Section~\ref{ss:hyper-parameter} of hyper-parameter study, we vary the embedding size for each method. 
Table~\ref{tab:overall_performance} shows the overall recommendation accuracy. We have the following main observations. 

\begin{table}[h]
	\begin{center}
		\caption{\textbf{Recommendation accuracy scores (\%)} of compared methods at embedding size 16.} \vspace{-5pt}
		\label{tab:overall_performance}
		\begin{tabular}{| l | c | c | c | c |} \hline
			& \multicolumn{2}{c|}{\textbf{MovieLens}} & \multicolumn{2}{c|}{\textbf{Pinterest}} \\ \hline 
			\textbf{Methods} & \textbf{HR} & \textbf{NDCG} & \textbf{HR} & \textbf{NDCG} \\ \hline
			\textbf{Pop} 	& 45.36 & 25.43 & 27.39 & 14.09 \\ \hline
			\textbf{ItemKNN}& 62.27 & 35.87 & 78.57 & 48.32 \\ \hline
			\textbf{MF-BPR}	& 66.64 & 39.73 & 86.90 & 54.01 \\ \hline
			\textbf{MF-eALS}& 67.88 & 39.83 & 87.13 & 52.55 \\ \hline
			\textbf{MLP}	& 68.41 & 41.03 & 86.48 & 53.85 \\ \hline
			\textbf{FISM} 	& 66.47 & 39.49 & 87.40 & 55.22 \\ \hline
			\textbf{NAIS-concat} & \textbf{69.72} & \textbf{41.96} & \textbf{88.44}	& \textbf{57.20}  \\ \hline
			\textbf{NAIS-prod} & \textbf{69.69} & \textbf{41.94} &	\textbf{88.44} & \textbf{57.22} \\ \hline
		\end{tabular}
	\end{center}
\end{table}

\begin{table*}[h]
	\begin{center}
		\caption{\textbf{Recommendation accuracy scores (\%)} of embedding-based methods at embedding size 8, 32, and 64. The best performance of each setting is highlighted as bold font.} \vspace{-5pt}
		\label{tab:overall_performance2}
		\begin{tabular}{| l | c | c | c | c | c | c | c | c | c | c | c | c |} \hline
			& \multicolumn{4}{c|}{\textbf{Embedding size = 8}} & \multicolumn{4}{c|}{\textbf{Embedding size = 32}} & \multicolumn{4}{c|}{\textbf{Embedding size = 64}} \\ \hline
			& \multicolumn{2}{c|}{\textbf{MovieLens}} & \multicolumn{2}{c|}{\textbf{Pinterest}} 	& \multicolumn{2}{c|}{\textbf{MovieLens}} & \multicolumn{2}{c|}{\textbf{Pinterest}} 	& \multicolumn{2}{c|}{\textbf{MovieLens}} & \multicolumn{2}{c|}{\textbf{Pinterest}} \\ \hline 
			\textbf{Methods} & \textbf{HR} & \textbf{NDCG} & \textbf{HR} & \textbf{NDCG}& \textbf{HR} & \textbf{NDCG} & \textbf{HR} & \textbf{NDCG}& \textbf{HR} & \textbf{NDCG} & \textbf{HR} & \textbf{NDCG} \\ \hline
			\textbf{MF-BPR}	& 62.86 & 36.08 & 85.85 & 53.26 & 68.54 & 41.14 & 86.34 & 54.54 & 68.97 & 41.91 & 85.8 & 54.58 \\ \hline
			\textbf{MF-eALS}& 62.8 & 36.35 & 86.26 & 51.86  & 70.4 & 42.16 & 86.75 & 53.84 & 70.35 & 43.5 & 85.77 & 53.77 \\ \hline
			\textbf{MLP}	& \textbf{67.1} & \textbf{39.98} & 85.9 & 53.67 & 69.24 & 42.51 & 86.77 & 54.2 & 70.18 & 42.64 & 86.9 & 54.5 \\ \hline
			\textbf{FISM} 	& 61.71 & 35.73 & 87.03 & 54.82 & 69.29 & 41.71 & 88.43 & 57.13 & 70.17 & 42.82 & 88.62 & 57.18 \\ \hline
			\textbf{NAIS-concat} & 64.17 & 37.36 & 87.44 & 55.27 & 70.83 & 43.36 & 88.56 & 57.47 & 71.66 & 44.15 & 88.74 & 57.75  \\ \hline
			\textbf{NAIS-prod} & 64.5 & 37.6 & \textbf{87.88} & \textbf{55.75} & \textbf{70.91} & \textbf{43.39} & \textbf{88.67} & \textbf{57.59} & \textbf{71.82} & \textbf{44.18} & \textbf{88.84} & \textbf{57.9}
			\\ \hline
		\end{tabular}
	\end{center}
	\vspace{-15pt}
\end{table*}

\begin{figure*}[t]
	\centering
	\begin{subfigure}[b]{0.235\textwidth}
		\centering
		\includegraphics[width=\textwidth]{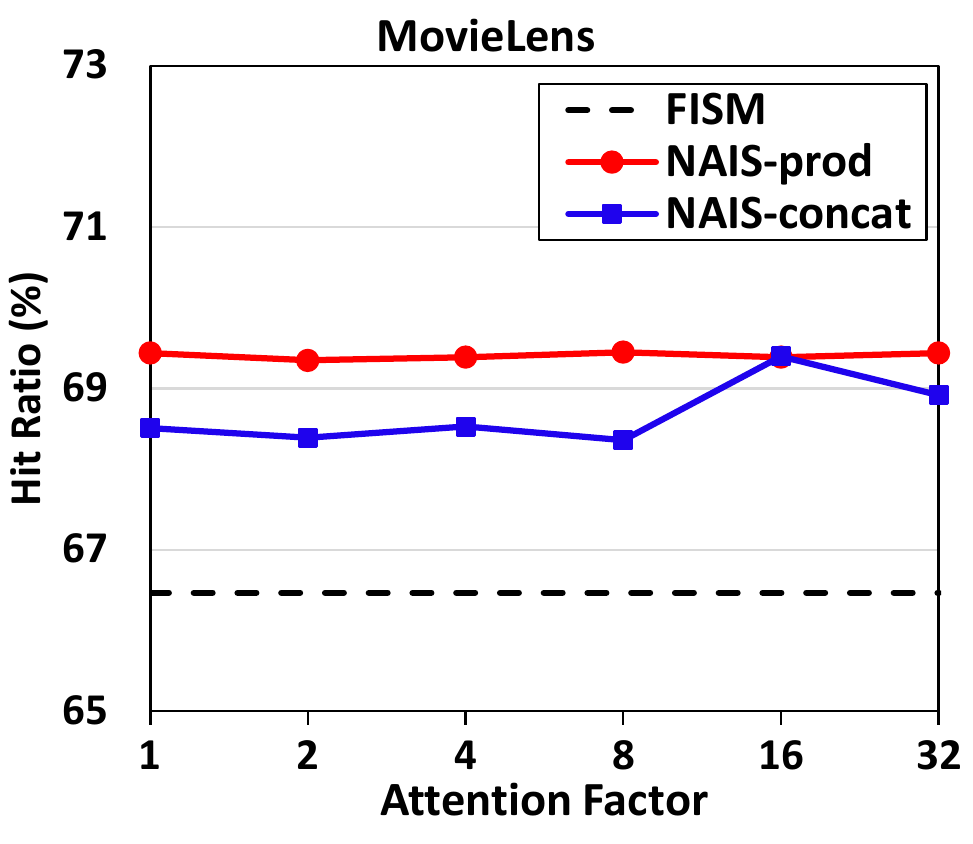}
		\vspace{-15pt}
		\caption{MovieLens --- HR}
		\label{fig:factor-ml-hr}
	\end{subfigure} 
	\begin{subfigure}[b]{0.235\textwidth}
		\centering
		\includegraphics[width=\textwidth]{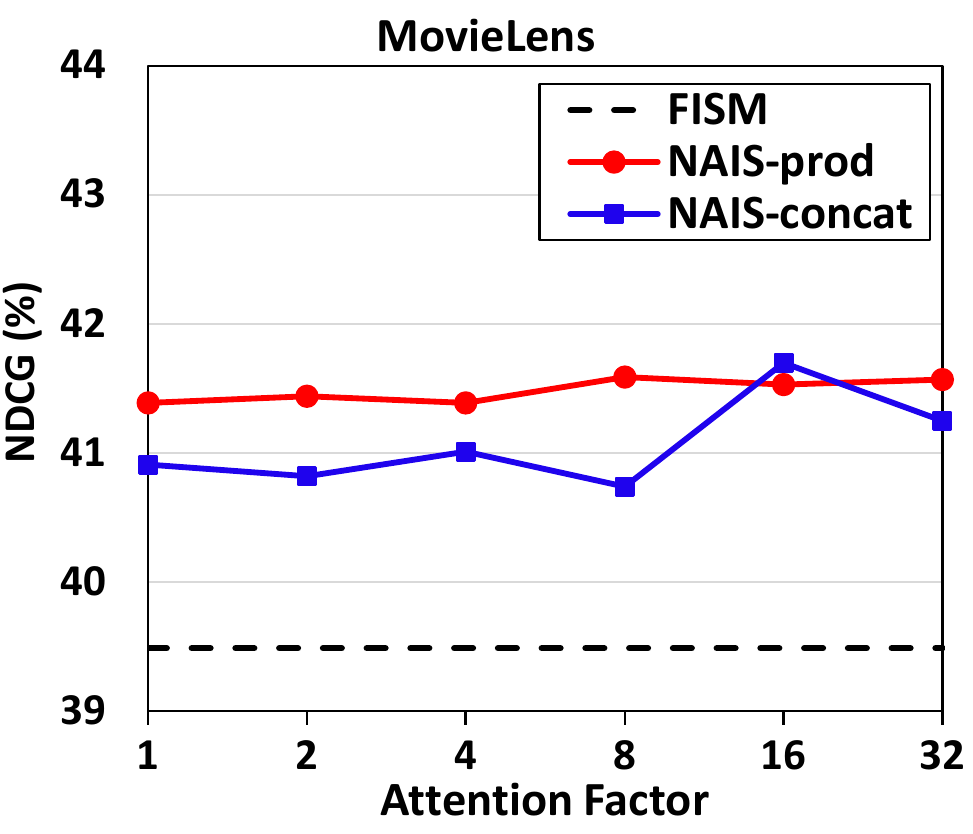}
		\vspace{-15pt}
		\caption{MovieLens --- NDCG}
		\label{fig:factor-ml-ndcg}
	\end{subfigure} 
	\begin{subfigure}[b]{0.235\textwidth}
		\centering
		\includegraphics[width=\textwidth]{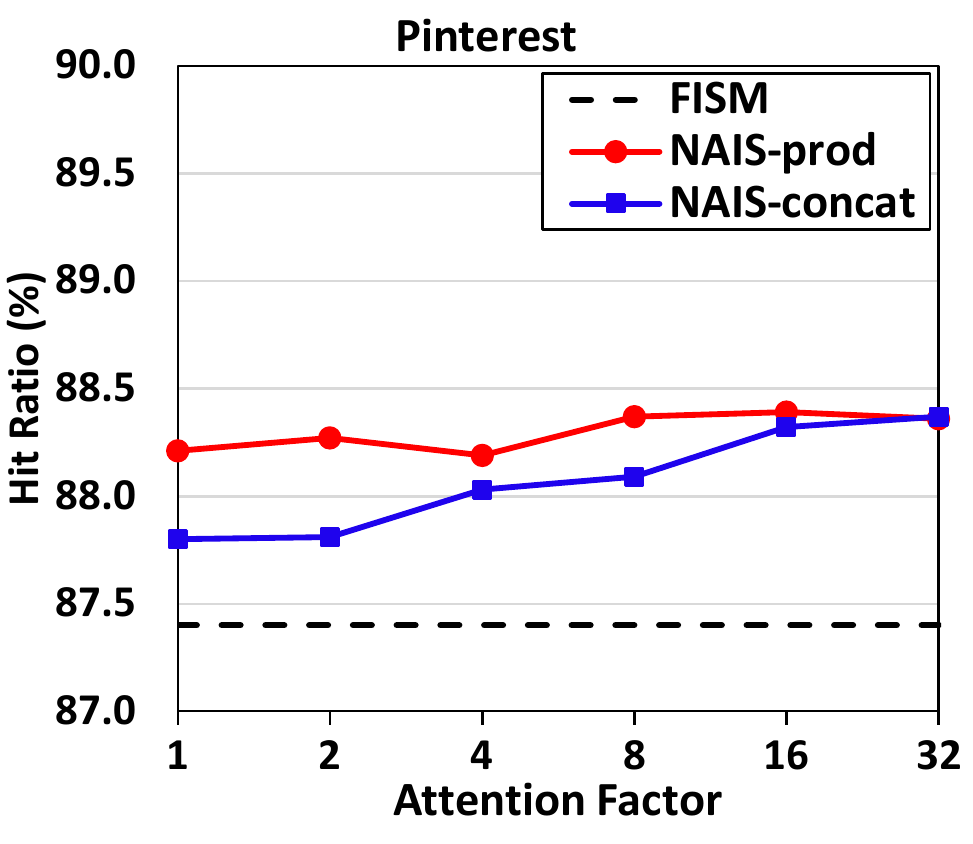}
		\vspace{-15pt}
		\caption{Pinterest --- HR}
		\label{fig:factor-pinterest-hr}
	\end{subfigure} 
	\begin{subfigure}[b]{0.235\textwidth}
		\centering
		\includegraphics[width=\textwidth]{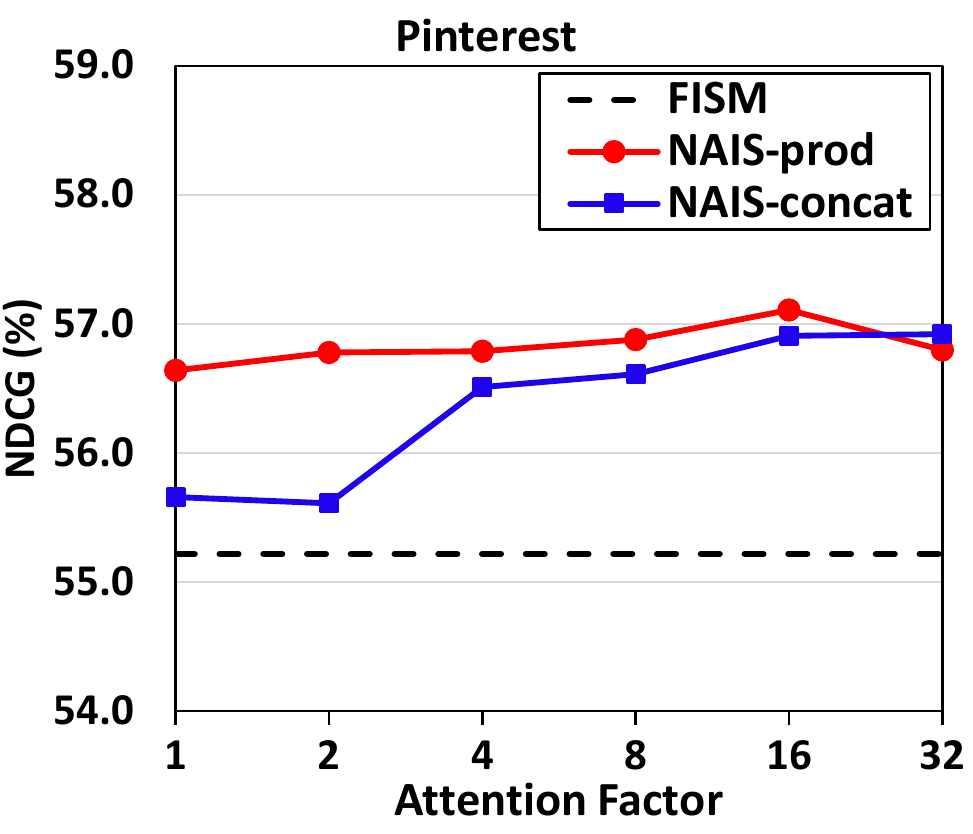}
		\vspace{-15pt}
		\caption{Pinterest --- NDCG}
		\label{fig:factor-pinterest-ndcg}
	\end{subfigure} 
	\caption{Testing performance of NAIS methods \wrt the attention factor $a$.}
	\vspace{-10pt}
	\label{fig:attention_factor}
\end{figure*}

\begin{figure*}[t]
	\centering
	\begin{subfigure}[b]{0.235\textwidth}
		\centering
		\includegraphics[width=\textwidth]{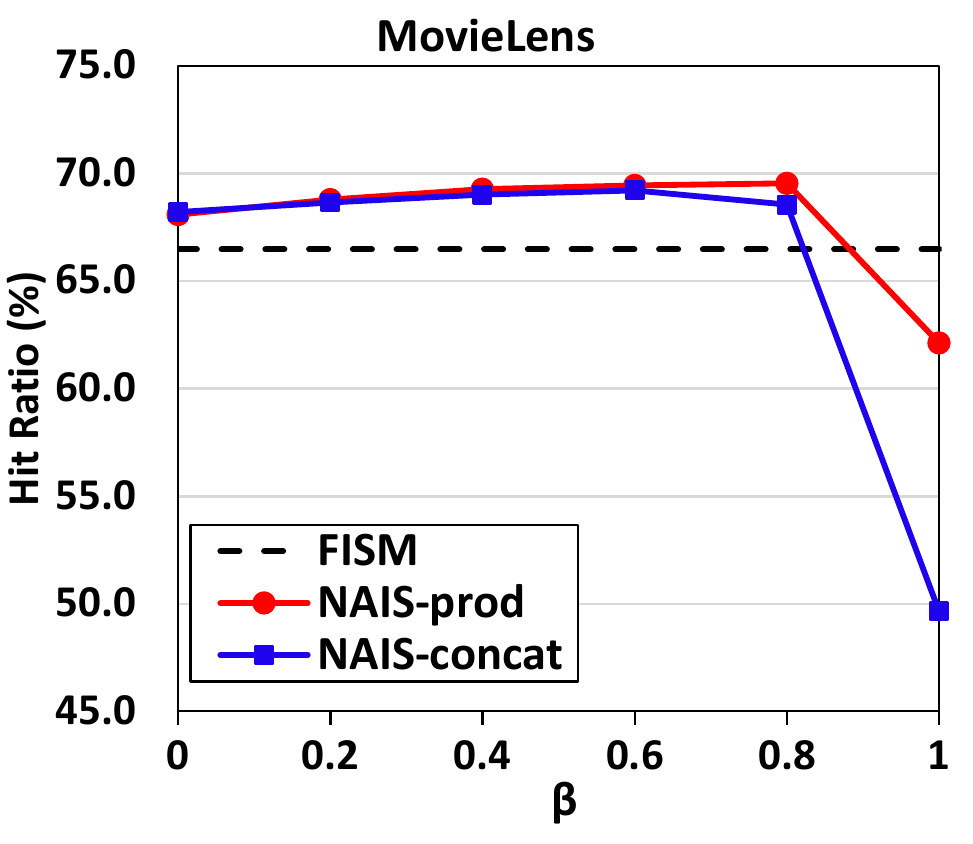}
		\vspace{-15pt}
		\caption{MovieLens --- HR}
		\label{fig:beta-ml-hr}
	\end{subfigure} 
	\begin{subfigure}[b]{0.235\textwidth}
		\centering
		\includegraphics[width=\textwidth]{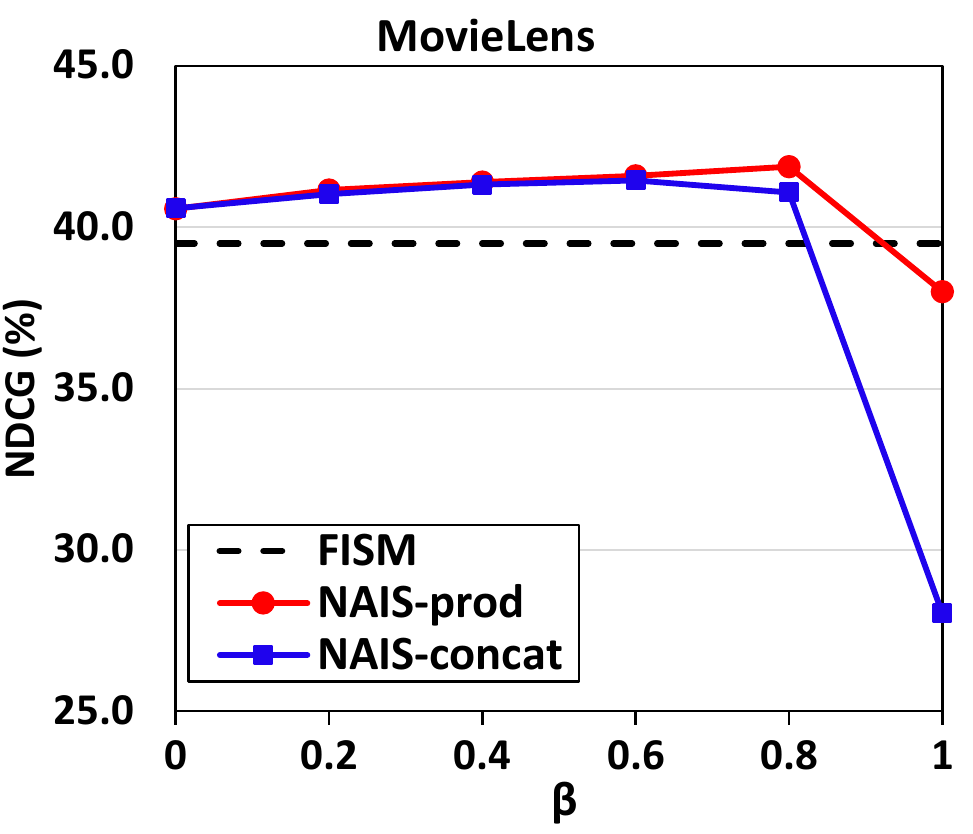}
		\vspace{-15pt}
		\caption{MovieLens --- NDCG}
		\label{fig:beta-ml-ndcg}
	\end{subfigure} 
	\begin{subfigure}[b]{0.235\textwidth}
		\centering
		\includegraphics[width=\textwidth]{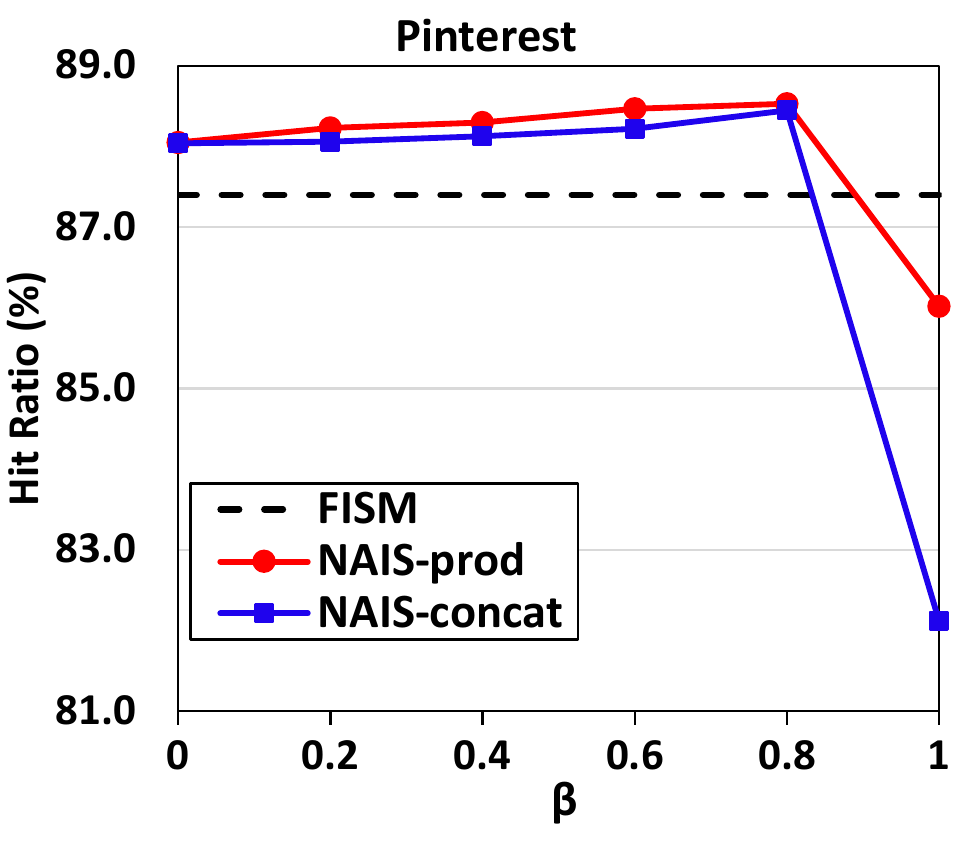}
		\vspace{-15pt}
		\caption{Pinterest --- HR}
		\label{fig:beta-pinterest-hr}
	\end{subfigure} 
	\begin{subfigure}[b]{0.235\textwidth}
		\centering
		\includegraphics[width=\textwidth]{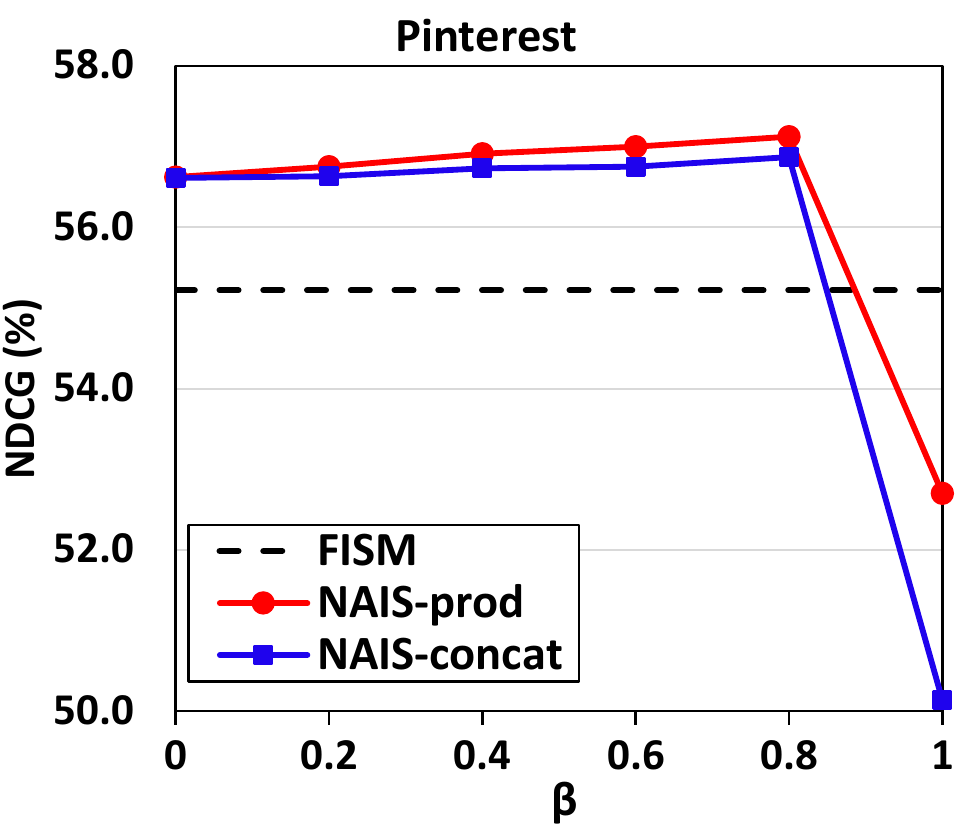}
		\vspace{-15pt}
		\caption{Pinterest --- NDCG}
		\label{fig:beta-pinterest-ndcg}
	\end{subfigure} 
	\caption{Testing performance of NAIS methods \wrt the smoothing exponent $\beta$.}
	\vspace{-10pt}
	\label{fig:beta}
\end{figure*}

\begin{itemize}
	\item 1. The two NAIS methods achieve the highest NDCG and HR scores on both datasets. They reach the same performance level, achieving significant improvements over other methods ($p<10^{-3}$ judged by the one-sample paired t-test). We believe the benefits are credited to the effective design of the attention networks in learning item-to-item interactions. 
	\item 2. Learning-based CF approaches perform better than heuristic-based approaches Pop and ItemKNN. In particular, FISM outperforms its counterpart ItemKNN with about $10\%$ relative improvements. Considering that both methods use the same prediction model while differ in the way of estimating item similarities, we can clearly see the positive effect of tailored optimization for recommendation. 
	\item 3. Among the baselines, there is no obvious winner between user-based CF models (MF, MLP) and item-based CF model (FISM). Specifically, on MovieLens user-based models perform better than FISM, while on Pinterest FISM outperforms user-based models. Since user interactions of the Pinterest data are more sparse, it reveals that item-based CF might be more advantageous for sparse datasets, which is in consistent with the finding in previous work \cite{FISM}.
\end{itemize}

It is worth pointing out that the performance of NAIS reported in Table \ref{tab:overall_performance} uses the default settings of hyper-parameters (reported in Section~\ref{ss:settings}). Further improvements can be observed by tuning hyper-parameters, which will be explored in the next subsection. 

\subsection{Hyper-parameter Study (RQ3)}
\label{ss:hyper-parameter}

By introducing an attention network, NAIS has two additional hyper-parameters --- the hidden layer size of the attention network (\aka the attention factor $a$) and the smoothing exponent $\beta$. 
In addition, as an embedding-based model, the embedding size is another crucial hyper-parameter for NAIS. 
This subsection investigates the impact of the three hyper-parameters. 


Table~\ref{tab:overall_performance2} shows the performance of embedding-based methods at embedding size 8, 32, and 64. We can see that the performance trends are generally in consistent with the observations at embedding size 16 (elaborated in Section~\ref{ss:performance}). Our NAIS methods achieve the best performance in most cases, with the only exception of embedding size 8, where MLP performs the best. This is because when the embedding size is small, linear models are limited by the small embedding size, while non-linear models are easy to express stronger representation ability than linear models. 

Figure \ref{fig:attention_factor} shows the performance of NAIS \wrt attention factor. We can see that regardless of the setting of attention factor, both NAIS methods outperform FISM. Among the two methods, NAIS-prod performs better than NAIS-concat for small attention factors, demonstrating the positive effect of using $\textbf{p}_i\odot\textbf{q}_j$ of as input to the attention network for learning the weight of $\textbf{p}_i^T\textbf{q}_j$. Moreover, using a large attention factor for NAIS-concat can compensate the performance gap between NAIS-prod. This implies the utility of using an expressive model for learning the attention weights. 

Figure \ref{fig:beta} shows the performance of NAIS \wrt $\beta$. It is clear that when $\beta$ is smaller than 1, both NAIS methods show good performance and outperform FISM. However, when $\beta$ is set to 1, the performances of NAIS degrade significantly and are worse than FISM. Note that setting $\beta$ to 1 means using softmax to normalize the attention weights, a standard setting for neural attention networks~\cite{bahdanau2014neural,ijcai2017-afm,ACF}. Unfortunately, such a standard setting does not work well for CF datasets. We believe the reason is caused by the large variance of the length of user histories. Specifically, on MovieLens and Pinterest, the $(mean, variance)$ of user history's length are $(166,37145)$ and $(27,57)$, respectively. Such a large variance on the number of attentive components seldom happens in NLP and CV tasks that deal with sentences (\ie attention on words) and images (\ie attention on regions). This is a key insight of this work for employing attention networks on user behavior data, which to our knowledge has never been investigated before.

\section{Related Work}
\label{sec:related}
Early works on CF mostly deal with explicit feedback like user ratings, formulating it as a rating prediction task~\cite{ICF,SVD++}. The target is to minimize the error between observed ratings and the corresponding model predictions. For this regression-based CF task, MF --- a linear latent factor model --- is known to be the most effective approach. 
Its basic idea is to associate each user and item with a latent vector (\aka embedding), modeling their matching score as the inner product between their latent vectors. 
Many variants to MF have been proposed, such as SVD++~\cite{SVD++}, Localized MF~\cite{Yongfeng:WWW2013}, Hierarchical MF~\cite{Suhang:IJCAI2015}, Social-aware MF~\cite{zhaoTKDE2016}, and Cross-Platform MF~\cite{Cao:2017:TOIS}. The SVD++ model has demonstrated strong representation power in fitting ratings; in particular, it is reported to be the best single model in the Netflix challenge. In our view, this shall be creditable to its integration of user-based CF and item-based CF under the latent factor model. While in the original paper of SVD++~\cite{SVD++}, the authors claimed to enhance MF by incorporating implicit feedback, the modeling of implicit feedback part is essentially an item-based CF model.

Later research efforts on CF have shifted towards learning recommenders from implicit feedback~\cite{BPR,FISM,fastMF,iCD,NCF}.
By nature implicit feedback is a one-class data, where only users' interaction behaviors are recorded and their explicit preferences on items (\ie likes or dislikes) are unknown. 
Distinct from early CF methods that predict rating scores, the works on implicit feedback typically treat CF as a personalized ranking task, adopting a ranking-based evaluation protocol on top-K recommendations. 
It is obvious that evaluating a CF method with a ranking-based protocol is more convincing and practically valuable, since recommendation is naturally a top-K ranking task for many applications. 
Moreover, there is empirical evidence showing that a CF model of lower rating prediction error does not necessarily result in higher accuracy in top-K recommendation~\cite{Cremonesi:2010}. 

Technically speaking, the key difference between rating prediction methods and top-K recommendation methods is in the way of optimizing the CF model~\cite{fastMF}. Specifically, rating prediction methods often optimize a regression loss on observed data only, while top-K recommendation methods need to account for missing data (\aka negative feedback)~\cite{Cremonesi:2010}. As such, it is technically feasible to tailor a rating prediction CF method for implicit feedback by simply adjusting the objective function to optimize. 

To learn a recommender model from implicit feedback, two types of learning to rank (L2R) objective functions have been commonly applied: pointwise and pairwise. Pointwise L2R methods either optimize a regression-based squared loss~\cite{FISM,wang2016scalable} or classification-based log loss~\cite{NCF}, by either sampling negative feedback from missing data~\cite{Chen:2017} or treating all missing data as negative feedback~\cite{fastMF}. 
For linear CF models like MF and its variants (\eg factorization machines), there exist efficient coordinate descent algorithms that optimize squared loss over all missing data~\cite{fastMF,iCD}. However for complex non-linear CF models like neural networks, only SGD-based optimization methods are applicable, and sampling negative feedback from missing data is necessary for the sake of efficiency. 
Pairwise L2R methods consider a pair of a user's positive and (sampled) negative feedback, maximizing the margin of their predicted scores regardless of their exact values~\cite{BPR,zhaoIJCAI17}. The underlying assumption is that an observed interaction should be more likely of interest to the user than an unobserved feedback. A state-of-the-art work develops adversarial personalized ranking~\cite{APR}, which employs adversarial training on pairwise learning to enhance the robustness of recommender models and improve their generalization performance. 

In recent years, using deep neural networks (DNNs, \aka deep learning) for recommendation becomes increasingly popular. DNNs have strong ability to learn complex functions from data, being well known for extracting high-level features from low-level raw data, such as images and audios~\cite{representation_learning}. 
Existing works on DNNs for recommendation can be divided into two types: 1) using DNNs for feature extraction from auxiliary data, \eg images and texts~\cite{Geng:2015,Suhang:WWW2017}, and 2) using DNNs for learning the user-item scoring function~\cite{NCF,NFM,TEM}. Since we focus on CF that leverages user-item interactions only, the second type of work is more relevant to this work. In \cite{NCF}, the authors formulated a general NCF framework for performing CF with feed-forward neural networks and devised three user-based CF models. 
Later on NCF is extended to incorporate attributes and optimize a pairwise ranking loss~\cite{SilkRoad}. 
The neural factorization machine (NFM)~\cite{NFM} is proposed to model higher-order and non-linear interactions among features, which is suitable for information-rich recommendation scenario, such as attribute-based and context-aware recommendation. 
More recently, 
Wang~\etal~\cite{TEM} combines the strengths of embedding-based with tree-based models for explainable recommendation. 

The work that is most similar to ours is the \textit{Attentive Collaborative Filtering} (ACF)~\cite{ACF}, which develops an attention network for user-based CF. Our NAIS differs from ACF and all previous works by tailoring the attention network for item-based CF. We find that using the standard attention network does not work well on user interaction histories, due to the problematic softmax in dealing with the varying-length histories. To address this, we propose to smooth the denominator of the softmax function. This insight is particularly useful for developing attention network for sequential data that has a large variance its length, which to our knowledge has never been explored before. 

\section{Conclusion}
\label{sec:conclusion}

In this work, we developed neural network methods for item-to-item collaborative filtering. 
Our key argument is that the historical items of a user profile do not contribute equally to predict the user's preference on an item. 
To address this point, we first revisited the FISM method from the perspective of representation learning, and then devised several attention mechanisms step by step to enhance its representation ability. 
We found that the conventional design of neural attention network~\cite{bahdanau2014neural,ACF,ijcai2017-afm,Li:2017:NAS} did not work well for item-based CF, due to the large variances of the lengths of user histories. We proposed a simple yet effective variant of softmax to address the large variance issue on user behaviors. 
We conducted empirical studies to validate the effectiveness of our NAIS methods. Experimental results show that NAIS significantly outperforms FISM, achieving competitive performance for the item recommendation task. 

To our knowledge, this is the first work on designing neural network models for item-based CF, opening up new research possibilities for future developments of neural recommender models. 
In future, we are particularly interested in exploring deep architectures for NAIS methods. Currently, our design of NAIS considers the pairwise similarities, \ie second-order interactions between items only, due to the consideration of keeping the model's simpleness in online personalization. 
This is primarily for the practical concern of a recommendation method. 
For further improvements on the recommendation accuracy, it is natural to extend NAIS by placing fully connected layers or convolutional layers above the embedding layer, which has been shown to be helpful by modeling high-order and nonlinear features interactions~\cite{NFM}. 
Technically speaking, another interesting direction worth exploring is to combine deep neural networks with graph-based methods~\cite{hong2017user,POH}, which have their unique strengths and have also been widely used for ranking.
Moreover, we are interested in exploring the recent adversarial personalized ranking learning on item-based CF to investigate the possible performance improvements~\cite{APR}. 
Lastly, we will investigate the explainability of recommender systems, which is a promising direction recently \cite{Zhang:2014:EFM,TriRank,Ren:2017,TEM} and can be facilitated by introducing attention networks on item-based CF methods.

\section*{Acknowledgments}
{
	NExT research is supported by the National Research Foundation, Prime Minister's Office, Singapore under its IRC@SG Funding Initiative. This work is partly supported by the National key research and development program of China under Grant 2017YFB1401304, by the National Research Foundation Singapore under its AI Singapore Programme, Linksure Network Holding Pte Ltd and the Asia Big Data Association (award No.: AISG-100E-2018-002).
}

\bibliographystyle{IEEEtran}
\bibliography{reference}

\vspace{-10pt}
\begin{IEEEbiography}[{\includegraphics[width=1in,height=1.25in,clip,keepaspectratio]{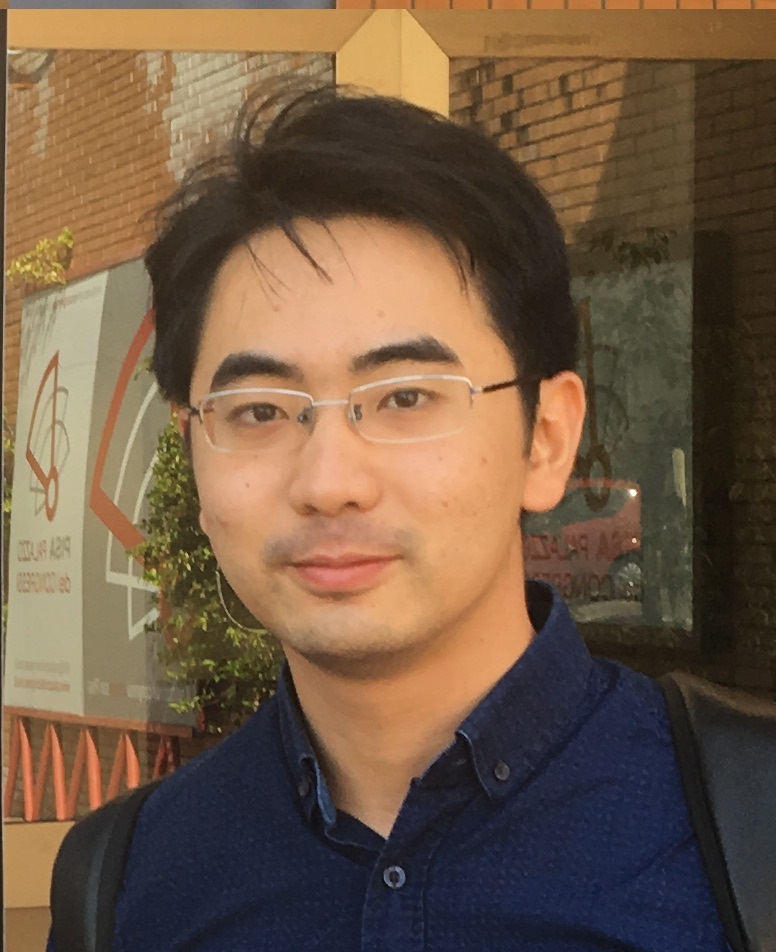}}]{Xiangnan He} 
	is currently a senior research fellow with School of Computing, National University of Singapore (NUS). He received his Ph.D. in Computer Science from NUS. His research interests span recommender system, information retrieval, and multi-media processing. He has over 20 publications appeared in several top conferences such as SIGIR, WWW, MM, CIKM, and IJCAI, and journals including TKDE, TOIS, and TMM. His work on recommender system has received the Best Paper Award Honorable Mention of ACM SIGIR 2016. Moreover, he has served as the PC member for the prestigious conferences including SIGIR, WWW, MM, AAAI, IJCAI, WSDM, CIKM and EMNLP, and the regular reviewer for prestigious journals including TKDE, TOIS, TKDD, TMM etc.
\end{IEEEbiography} \vspace{-10pt}
\begin{IEEEbiography}[{\includegraphics[width=1in,height=1.25in,clip,keepaspectratio]{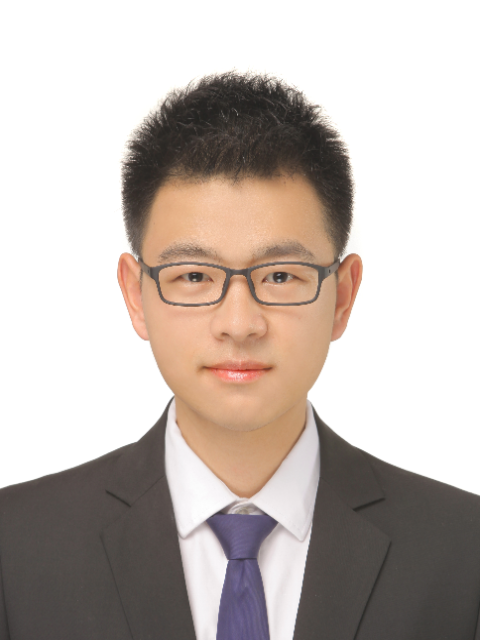}}]{Zhankui He} is an undergraduate student at Fudan University, China and an exchange student at National University of Singapore(NUS). He is a research assistant in Lab of Media Search and NExT research center in NUS. He has been awarded the 2016 Shanghai Scholarship, the 2016 Fudan Excellent Student Award and the 2017 Oriental CJ Scholarship. His research interest includes recommendation system and computer vision.
\end{IEEEbiography} \vspace{-10pt}
\begin{IEEEbiography}[{\includegraphics[width=1in,height=1.25in,clip,keepaspectratio]{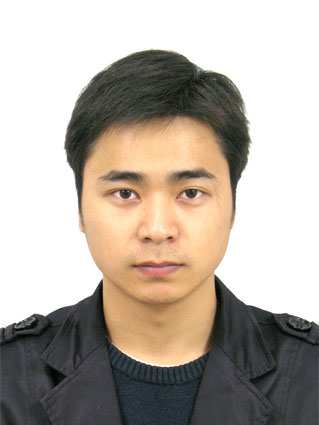}}]{Jingkuan Song} 
 is currently a Professor in University of Electronic Science and Technology of China. He joined Columbia University as a Postdoctoral Research Scientist (2016-2017), and University of Trento as a Research Fellow (2014-2016). He obtained his PhD degree in Information Technology from The University of Queensland (UQ), Australia, in 2014. His research interest includes large-scale multimedia retrieval, image/video segmentation and image/video annotation using hashing, graph learning and deep learning techniques.
\end{IEEEbiography}
\begin{IEEEbiography}[{\includegraphics[width=1in,height=1.25in,clip,keepaspectratio]{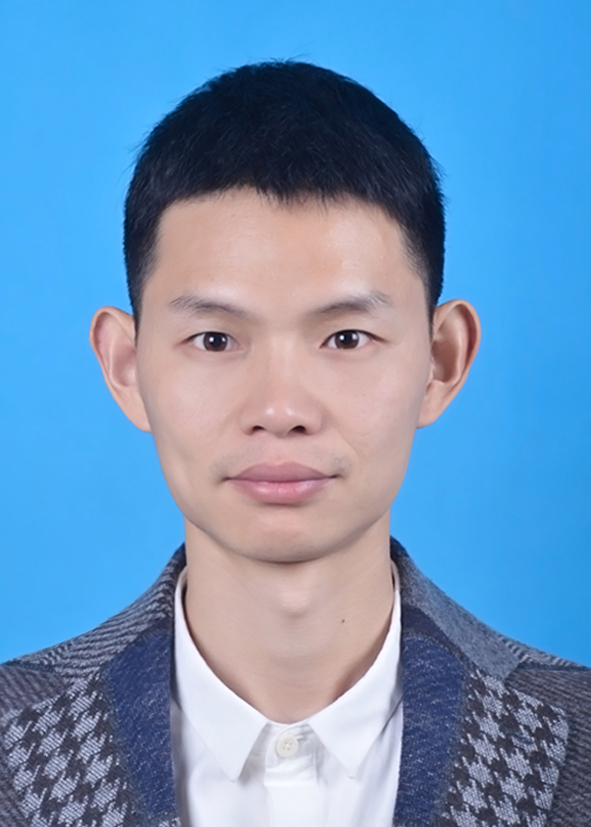}}]{Zhenguang Liu} 
	is currently a research fellow in Singapopre Agency for Science, Technology and Research (A* STAR). He was a research fellow in National University of Singapore from 2015 to May 2017. He respectively received his Ph.D. and B.E. degrees from Zhejiang University and Shandong University, China, in 2010 and 2015. His research interests include multimedia data analysis and data mining. Various parts of his work have been published in first-tier venues including TIP, AAAI, MM, TMM, TOMM. Dr. Liu has served as technical program committee member for conferences such as ACM MM and MMM, and the reviewer for IEEE Transactions on visualization and computer graphics, ACM MM, IEEE Transactions on Multimedia, Multimedia Tools and Applications, etc.
\end{IEEEbiography} 
\begin{IEEEbiography}[{\includegraphics[width=1in,height=1.25in,clip,keepaspectratio]{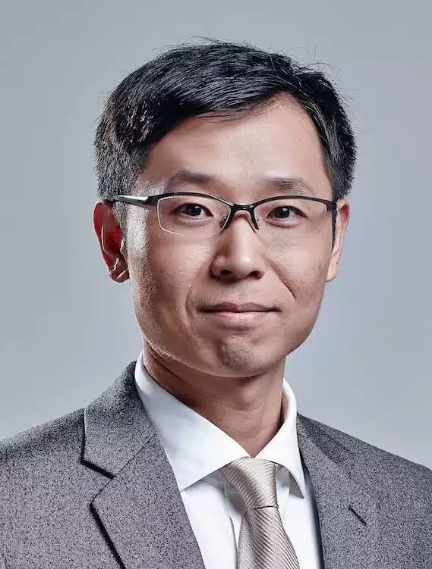}}]{Yu-Gang Jiang} 
	is Professor of Computer Science and Director of Shanghai Engineering Research Center for Video Technology and System at Fudan University, China. His Lab for Big Video Data Analytics conducts research on all aspects of extracting high-level information from big video data, such as video event recognition, object/scene recognition and large-scale visual search. He is the lead architect of a few best-performing video analytic systems in worldwide competitions such as the annual U.S. NIST TRECVID evaluation. His work has led to many awards, including the inaugural ACM China Rising Star Award, the 2015 ACM SIGMM Rising Star Award, and the research award for outstanding young researchers from NSF China. He was also selected into China's National Ten-Thousand Talents Program and the Chang Jiang Scholars Program of China Ministry of Education. He is currently an associate editor of ACM TOMM, Machine Vision and Applications (MVA) and Neurocomputing. He holds a PhD in Computer Science from City University of Hong Kong and spent three years working at Columbia University before joining Fudan in 2011.
\end{IEEEbiography} 
\begin{IEEEbiography}[{\includegraphics[width=1in,height=1.25in,clip,keepaspectratio]{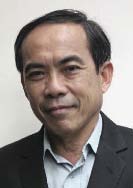}}]{Tat-Seng Chua} 
	is the KITHCT Chair Professor at the School of Computing, National University of Singapore. He was the Acting and Founding Dean of the School from 1998-2000. Dr Chua’s main research interest is in multimedia information retrieval and social media analytics. In particular, his research focuses on the extraction, retrieval and question-answering (QA) of text and rich media arising from the Web and multiple social networks. He is the co-Director of NExT, a joint Center between NUS and Tsinghua University to develop technologies for live social media search. Dr Chua is the 2015 winner of the prestigious ACM SIGMM award for Outstanding Technical Contributions to Multimedia Computing, Communications and Applications. He is the Chair of steering committee of ACM International Conference on Multimedia Retrieval (ICMR) and Multimedia Modeling (MMM) conference series. Dr Chua is also the General Co-Chair of ACM Multimedia 2005, ACM CIVR (now ACM ICMR) 2005, ACM SIGIR 2008, and ACMWeb Science 2015. He serves in the editorial boards of four international journals. Dr. Chua is the co-Founder of two technology startup companies in Singapore. He holds a PhD from the University of Leeds, UK.
\end{IEEEbiography}
\vfill
\end{document}